\documentclass[letterpaper, paper,11pt]{AAS}	

\usepackage{bm}
\usepackage{amsmath}
\usepackage{gensymb}
\usepackage{subfigure}
\usepackage{flafter}
\usepackage{placeins}
\usepackage{longtable,tabularx}

\usepackage{mdframed}
\usepackage[colorlinks=true, pdfstartview=FitV, linkcolor=black, citecolor= black, urlcolor= black]{hyperref}
\usepackage{overcite}
\usepackage{footnpag}

\usepackage{graphicx}
\usepackage{empheq}
\usepackage{caption}

\usepackage{comment}
\usepackage{url}
\usepackage{amssymb}
\usepackage{enumitem}
\usepackage{empheq}
\usepackage{float}
\usepackage{stackengine}

\newlength\dlf  

\newcommand{\trunc}{\text{Trunc}}
\newcommand{\forceindent}{\leavevmode{\parindent=1em\indent}}

\PaperNumber{21-222}

\begin{document}
\setlength{\abovedisplayskip}{.4em}
\setlength{\belowdisplayskip}{.4em}
\setlength{\abovedisplayshortskip}{.4em}
\setlength{\belowdisplayshortskip}{.4em}
\setlength{\textfloatsep}{5pt}
\title{A Rapid Method For Orbital Coverage Statistics With $\mathbf{J_2}$ Using Ergodic Theory}

\author{Andrew J. {Graven\thanks{Department of Mathematics, Cornell University, Ithaca, NY, 14853, United States}},\; Alan H. Barr\thanks{Department of Computing and Mathematical Sciences, California Institute of Technology, Pasadena, CA, 91125,  United States},\; and\; Martin W. Lo\thanks{Principal Engineer, Mission Design and Navigation Section, Jet Propulsion Laboratory, California Institute of Technology, Pasadena, CA, 91109, United States}}

\maketitle{} 		
\begin{abstract}
Quantifying long-term statistical properties of satellite trajectories typically entails time-consuming trajectory propagation. We present a fast, ergodic\cite{Arnold} method of analytically estimating these for $J_2-$perturbed elliptical orbits, broadly agreeing with trajectory propagation-derived results. We extend the approach in Graven and Lo (2019)\cite{GravenLo2019} to estimate: (1) Satellite-ground station coverage with limited satellite field of view and ground station elevation angle with numerically optimized formulae, and (2) long-term averages of general functions of satellite position. This method is fast enough to facilitate real-time, interactive tools for satellite constellation and network design, with an approximate $1000\times$ GPU speedup.
\end{abstract}
\begin{figure}[htb]
\centering
\subfigure[Space tracks of a satellite orbit (red at apogee and yellow at perigee) intersecting the visibility cones (blue) of two ground stations (white). Computing the portion of time the satellite spends in view of its ground station via trajectory propagation is relatively slow. However, using ergodic theory, we're able to rapidly estimate this quantity by integrating over the interior of the cone with respect to a carefully chosen probability measure. See \textbf{Figure \ref{fig:GSMaskGeometry}} for a definition of the cone.]{
\centering
\includegraphics[width=.46\textwidth]{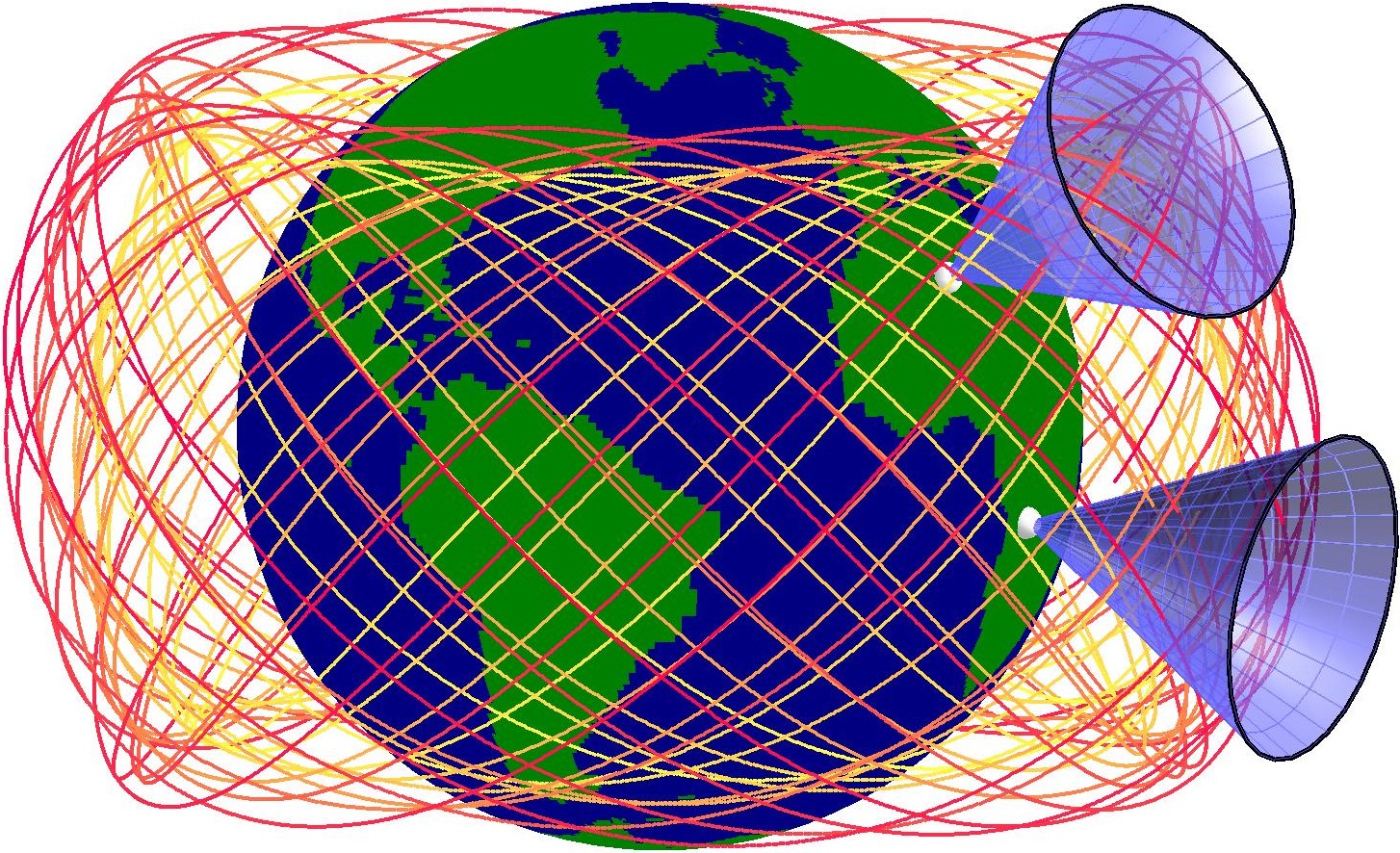}
\label{fig:GSConeVis}
}\hspace{2mm}
\subfigure[An example from a key class of applications: real time data generation \& visualization. This is a 3-dimensional heat map of percent visibility, from $\rho$, defined in Equation \ref{eqn:VPRho} and computed using Equation \ref{eqn:EllipticalVPRFormula}, for $64,000$ distinct satellite orbits, generated in 4.9 seconds on a dual core 2.7GHz laptop CPU, or 2 milliseconds on the Titan V GPU. The plot represents $64,000$ orbits generated from a $40\times40\times40$ grid of semimajor axis, eccentricity and inclination values.]{
\centering
\includegraphics[width=.47\textwidth]{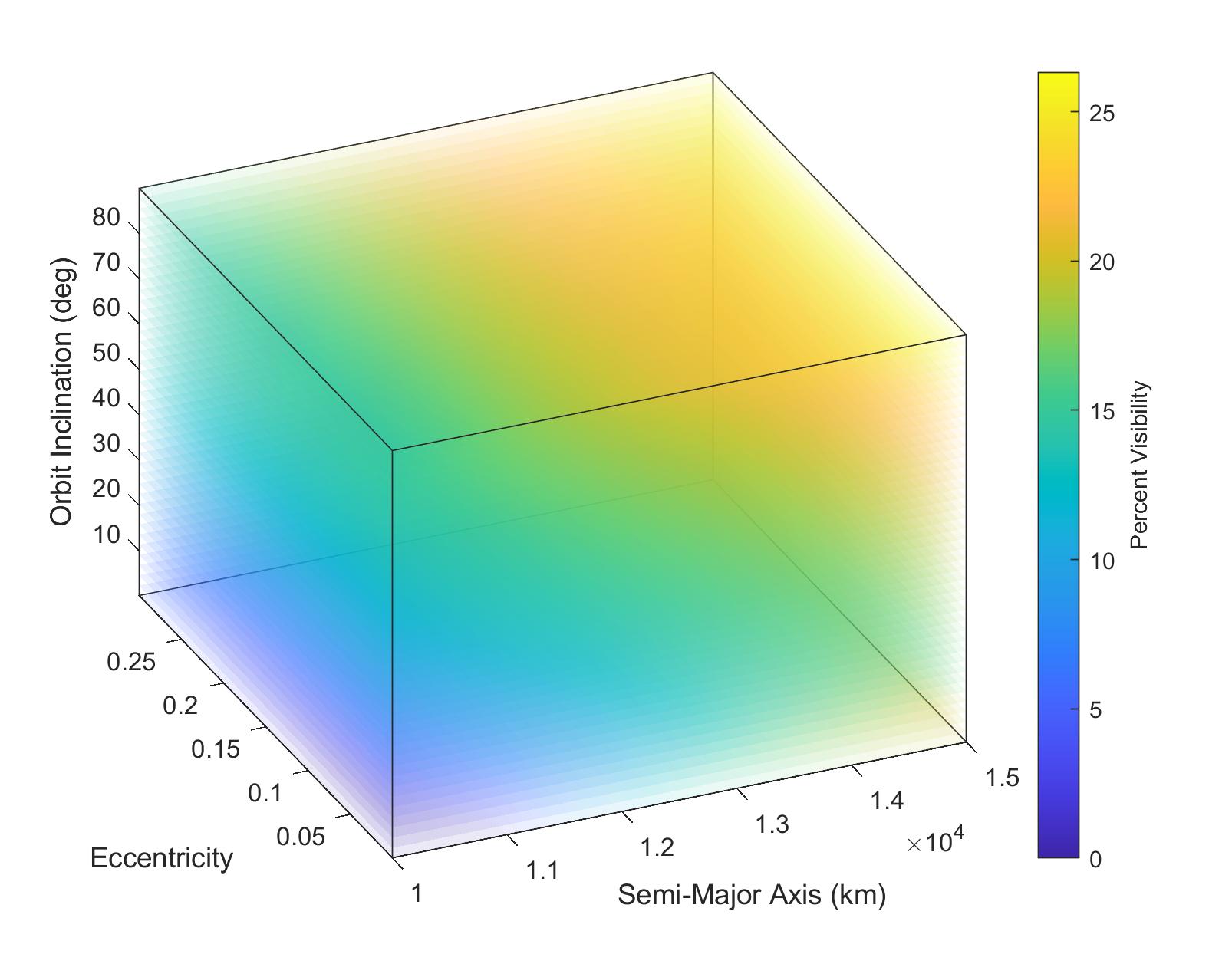}
\label{fig:VisHeatMap}
}\vspace{-.2em}
\refstepcounter{figure}\label{fig:IntroVis}
\vspace{-1em}
\end{figure}

\section{Introduction}\vspace{-.2em}
Dynamical Systems theory can be divided into two areas; the first consists of methods of solving differential equations and is well known. The second area is ergodic theory, which is concerned with the coverage and transport properties of dynamical systems and is less well known, perhaps due to its difficulty. Poincar\'e made fundamental contributions to both areas, creating the geometric theory of differential equations and pioneering the study of deterministic chaos in dynamical systems. Examples of problems and results in the domain of ergodic theory are the Poincar\'e recurrence theorem, the ergodic hypothesis of statistical mechanics, and the coverage of ground stations by satellites in orbit about a central body. This last example is the primary focus of this paper.

Advanced satellite and satellite constellation planning often requires an understanding of the long-term behavior of the proposed orbit or constellation. Quantities such as average satellite-ground station visibility, atmospheric drag and sun $\beta-$angle may inform the likelihood that mission requirements are satisfied. Estimating these by trajectory propagation can be costly and time consuming due to the time scales and small step sizes necessary for accurate estimates. This is compounded by the high-dimensionality of the design space, adding a combinatorial challenge finding acceptable (or, what's more, optimal) designs. Thus in many cases, such an approach may require a sparse sampling of the design space, the use of low-fidelity simulations, or significant computational resources.\vspace{-.2em}

In this paper we present a fast analytical approach to estimating a wide range of long-term statistics of aperiodic $J_2$-perturbed circular and elliptical orbits. We provide formulae for estimating: (1) Satellite-ground station view period ratios, with limited satellite field of view (FOV) and ground station elevation angle taken into account. (2) Averages of general functions of satellite position: drag force, gravity gradient, radiation, etc. Applying the Birkhoff-Kinchin Theorem of Ergodic theory, we express these quantities in terms of a definite integral. In certain cases, symmetries may be exploited to reduce the dimension of the integral, thereby further accelerating numerical evaluation. The evaluation of these formulae is sufficiently fast that it's feasible to use these in real-time concurrent engineering applications. These formulae turn out to be independent of the value of $J_2$ as long as $J_2\neq0$. Thus the results presented in this paper can be applied to any body with $J_2\neq0$. The value of $J_2$ is only necessary to verify the aperiodicity of the orbit to guarantee its ergodicity.

The mathematical symbols used throughout the paper are explained in-line and collected in the ``Notation'' section at the end of the paper. For the sake of compactness of notation, we define a symmetric truncation function, $\trunc_b(x)$ sending $x$ to the nearest point in $[-b,b]$:
\begin{align*}
    \trunc_b(x)&:=\min\{\max\{x,-b\},b\}\\
\intertext{Throughout the paper it will also be convenient to extend the domain of $\cos^{-1}$ and $\sin^{-1}$:}
    \sin^{-1}(x)&:=\sin^{-1}(\trunc_1(x))\\
    \cos^{-1}(x)&:=\cos^{-1}(\trunc_1(x))
\end{align*}
\begin{figure*}[htb]
\centering
\stackunder{\includegraphics[width=.31\textwidth]{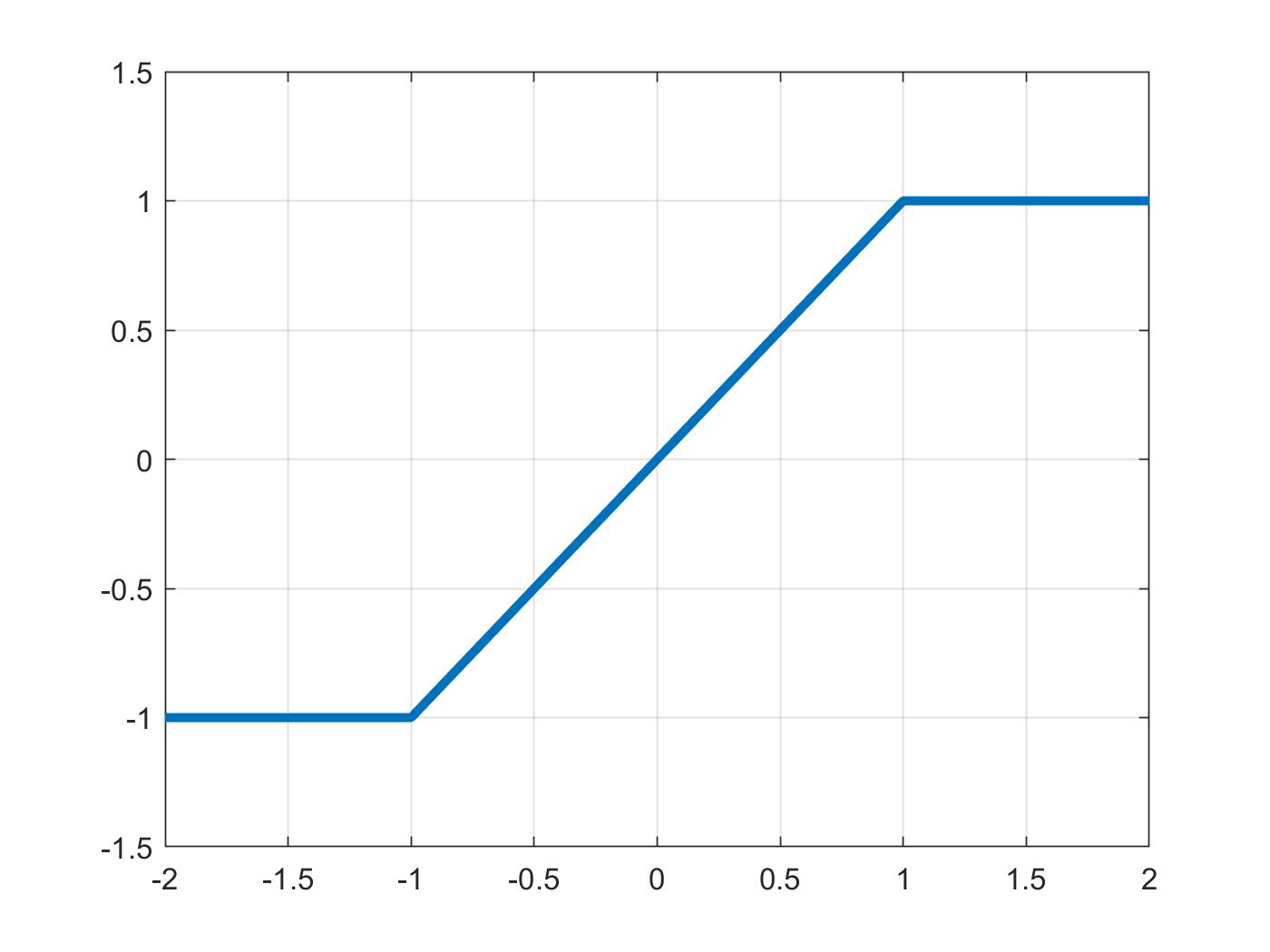}}{$\trunc_1(x)$}\hspace{1mm}
\stackunder{\includegraphics[width=.31\textwidth]{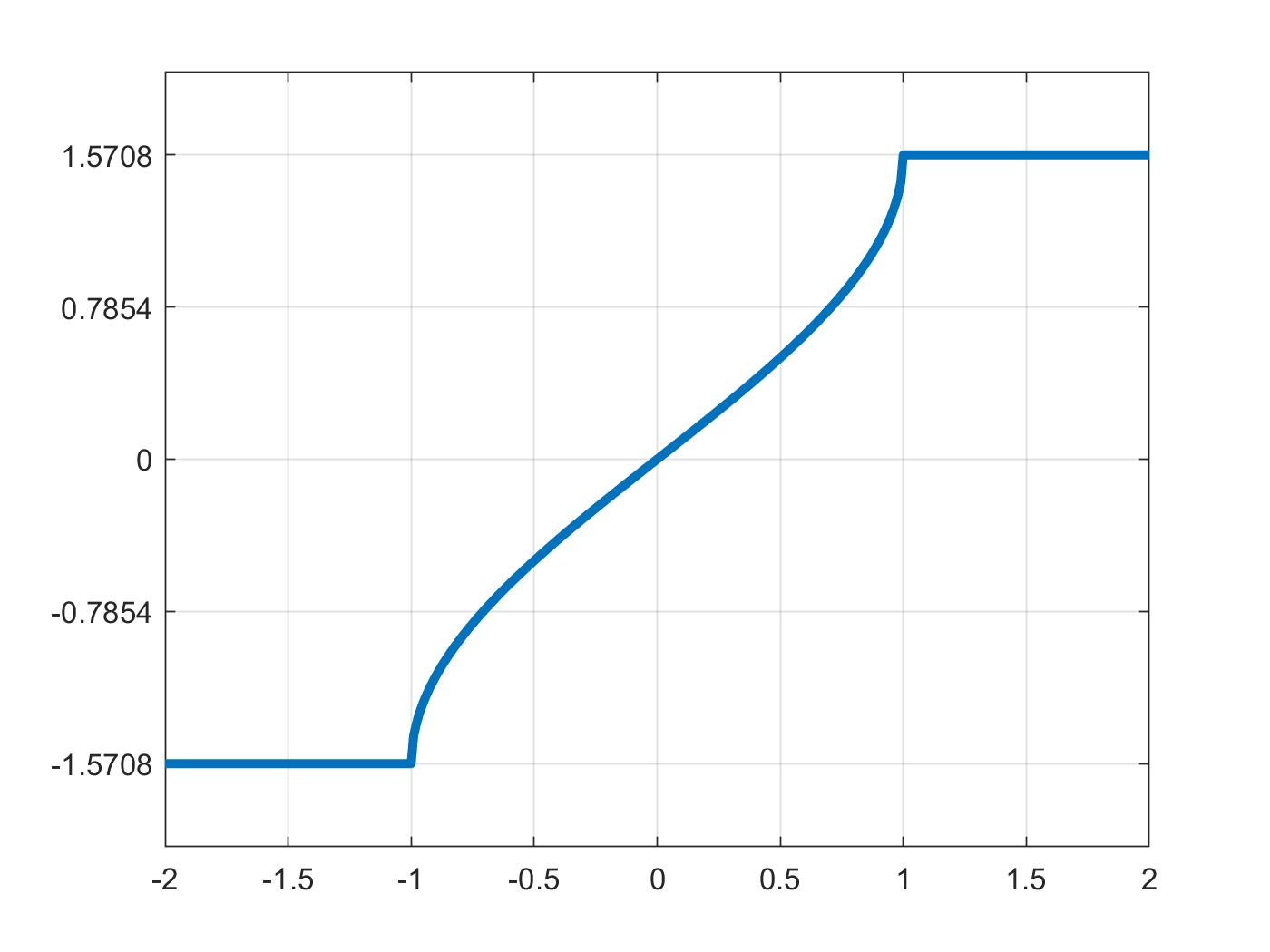}}{$\sin^{-1}(x)$}\hspace{1mm}
\stackunder{\includegraphics[width=.31\textwidth]{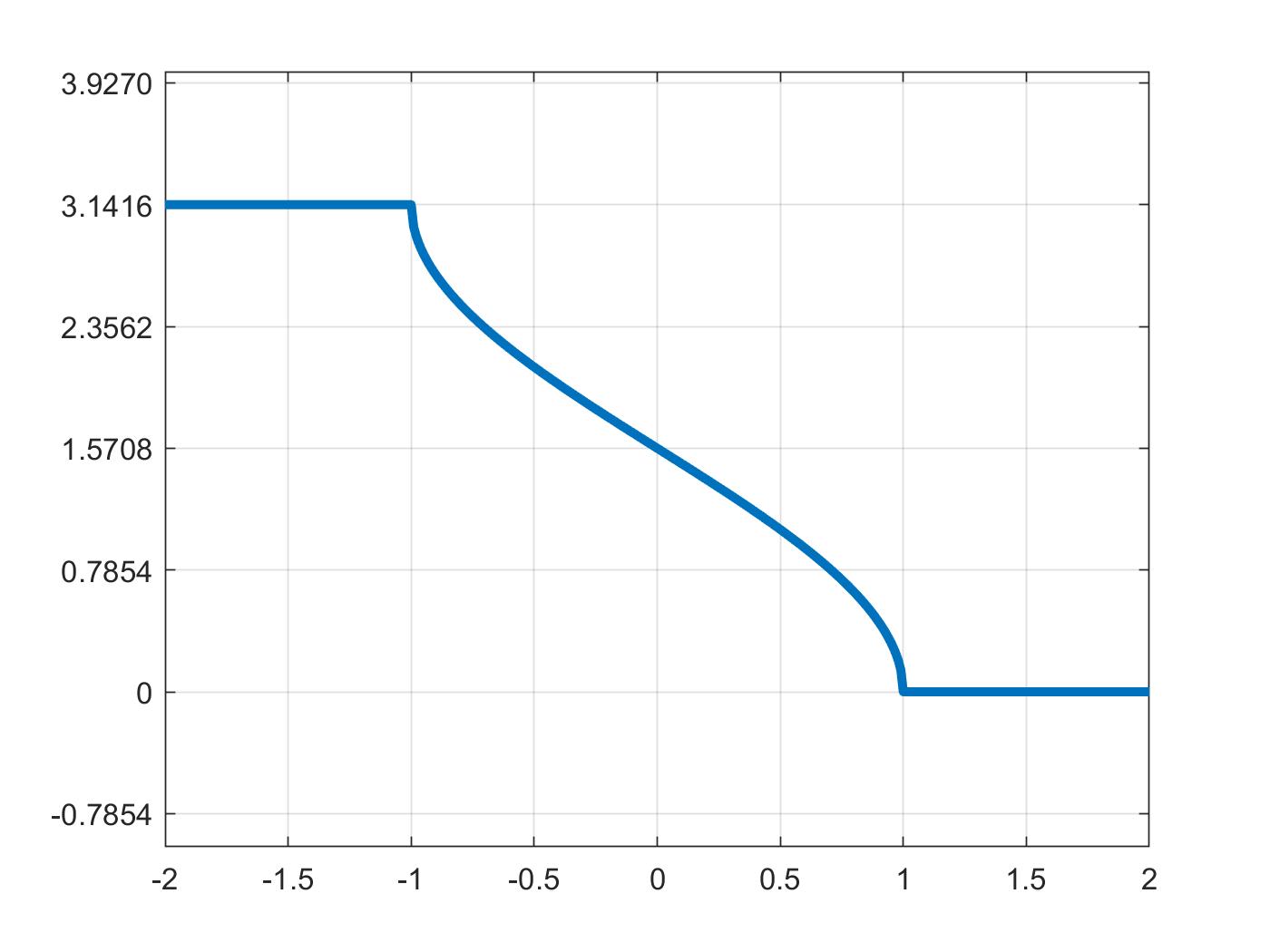}}{$\cos^{-1}(x)$}
\end{figure*}

\subsection[The View Period Ratio $\rho$]{The View Period Ratio \boldmath$\rho$}\vspace{-.5em}
The view period ratio, $\rho$ (also referred to as the ``view period percentage'') for a given satellite-ground station pair is the asymptotic fraction of time the satellite is able to communicate with its ground station. Specifically, if $T$ is the total flight time, and $P(T)$ is the portion of flight time for which the satellite and the ground station are able to communicate, then the view period ratio, $\rho$, is given by the limit:
\begin{equation}
    \rho=\lim\limits_{T\to\infty}\dfrac{P(T)}{T}
    \label{eqn:VPRho}
\end{equation}

The existence of this limit implies that the approximation $\rho T\approx P(T)$ improves\footnote{In terms of relative error, not necessarily absolute error.} as $T$ grows. And, under very mild assumptions, one can show that this limit indeed exists. Rapid computation of this quantity in various contexts is a central focus of this paper.

\subsection[The $J_2$ Model]{The \boldmath$J_2$ Model}
As a consequence of the Earth's rotation, it isn't a perfect sphere, but an oblate spheroid. This breaks the assumption of radial symmetry of the central body in the standard 2 Body Problem, resulting in a perturbed potential and equations of motion. This perturbation can be quantified with a spherical harmonic model of the central body, which allows us to describe the perturbed body by a sequence of coefficients $J_2,J_3,...$, the normalized zonal harmonic gravitational coefficients. For many applications, the $J_2$ term effectively dominates the others. The first two terms for Earth are: $J_2=1.083\cdot10^{-3}$ and $J_3=-2.5\cdot10^{-6}$, with magnitude continuing to drop off for higher order harmonics. Thus, most of this perturbation can be captured by the effect of $J_2$ alone. The $J_2-$perturbed equations of motion are given in Equation \ref{eqn:NonlinMotionEqs}.
\begin{equation}
\begin{aligned}
\label{eqn:NonlinMotionEqs}
\ddot{x}&=-\dfrac{\mu x}{r^3}(1-J_2\dfrac{3R_B^2}{2r^2}(\dfrac{5z^2}{r^2}-1))\\
\ddot{y}&=-\dfrac{\mu y}{r^3}(1-J_2\dfrac{3R_B^2}{2r^2}(\dfrac{5z^2}{r^2}-1))\\
\ddot{z}&=-\dfrac{\mu z}{r^3}(1-J_2\dfrac{3R_B^3}{2r^3}(\dfrac{5z^2}{r^2}-3))\\
r&=(x^2+y^2+z^2)^\frac{1}{2}
\end{aligned}
\end{equation}

\subsection[The Mean Linear $J_2$ Model]{The Mean Linear \boldmath$J_2$ Model}
Vallado\cite{Vallado} shows that the average secular motion of this system is well-approximated by a linear precession of the orbital elements: $M=$ mean anomaly, $\Omega=$ Longitude of the Ascending Node, $\omega=$ Argument of Periapsis. In particular, much of the long-term behavior of the system is accurately captured by the linear flow:
\begin{equation}
\begin{aligned}
\label{LinOrbEls}
M(t)&=M_0+\dot{M}t\\
\Omega(t)&=\Omega_0+\dot{\Omega}t \text{, mod }2\pi\\
\omega(t)&=\omega_0+\dot{\omega}t \text{, mod }2\pi
\end{aligned}
\end{equation}
Withs rates:
\begin{equation}
\begin{aligned}
\label{LinOrbElConsts}
\dot{M} &= \sqrt{\dfrac{\mu}{a^3}}[1+3J_2\dfrac{R_B^2}{4a(1-e^2)^{\frac{3}{2}}}(3\cos^2(i)-1)]\\
\dot{\Omega} &= -\sqrt{\dfrac{\mu}{a^3}}\dfrac{3}{2}\dfrac{R_B^2}{a^{2}(1-e^2)^2}J_2\cos(i)+\Omega_B\\
\dot{\omega} &= \sqrt{\dfrac{\mu}{a^3}}\dfrac{3}{4}\dfrac{R_B^2}{a^{2}(1-e^2)^2}J_2(4-5\sin^2(i))
\end{aligned}
\end{equation}
This model is used throughout the paper.

\subsection{The Invariant Measure \boldmath$\mu$}\vspace{-.5em}
A key result in Ergodic theory, the Birkhoff–Khinchin Theorem, asserts a time mean-space mean equivelence for a certain class of  ``Ergodic'' dynamical systems, see Arnold 1989 \cite{Arnold} and Sinai 1976 \cite{Sinai}. If $\phi_t(x):S\times\mathbb{R}\rightarrow S$ is the trajectory of an an ergodic system starting at $x\in S$, then the Birkhoff-Kinchin Theorem asserts that, for almost every $x\in S$, there exists a probability measure $\mu$ such that for any measurable function, $f:S\rightarrow\mathbb{R}$:
\begin{equation}
\label{eqn:BirkhoffKinchin}
\lim_{T\to\infty}\dfrac{1}{T}\int\limits_{t=0}^Tf(\phi_t(x))dt=\int\limits_Sf(x)d\mu
\end{equation}\vspace{-.1em}
The probability measure, $\mu$, (also referred to an an invariant\footnote{invariant in the sense that if we define the flow of the system: $F:S\times\mathbb{R}_+\rightarrow S$ by $F(x,t)=\vec{x}(t)$ s.t. $\vec{x}(0)=x$, then $\forall U\subseteq S,\;\forall t>0,\;\;\mu(F(U,t))=\mu(U)$} measure) can be interpreted as the infinitesimal proportion of time the state of the system spends at any given point in its state space. For example, if $\mu$ were the uniform distribution: $\mu=\text{Vol}(S)^{-1}$, we could conclude that in the long-term, $\vec{x}(t)$ spends the same amount of time in each region of its state-space.

\begin{figure}[htb]
\centering
\subfigure[Ground tracks of a circular orbit. The tracks exhibit a higher density at extreme latitudes, which is in agreement the well-known fact that ground stations at higher latitudes tend to have a higher level of connectivity with their satellites. Note that the long-term density of the ground tracks is longitude-independent.]{
\centering
\includegraphics[width=.464\textwidth]{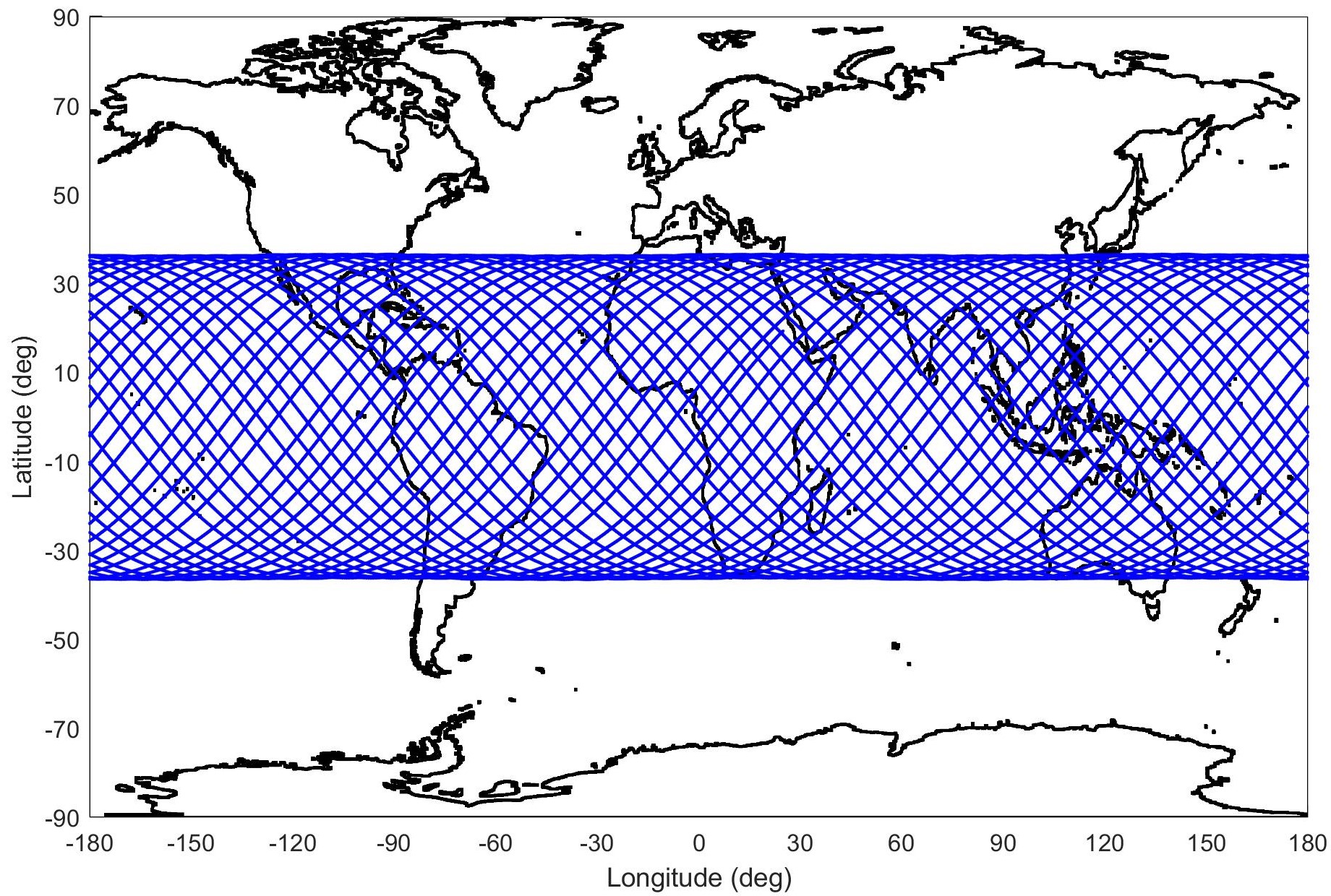}
\label{fig:GroundTrackDensity-a}
}\hspace{2mm}
\subfigure[Normalized${}^\ast$ probability density $\mu_{\text{c}}$ of circular satellite orbit position, from Equation \ref{eqn:CircularMeasure}. $\mu_{\text{c}}$ is a measure of the infinitesimal fraction of time the satellite spends at any given point in its state-space. Darker blue indicates a higher probability. The probability density closely corresponds with the distribution of ground tracks in \subref{fig:GroundTrackDensity-a}.]{
\centering
\includegraphics[width=.487\textwidth]{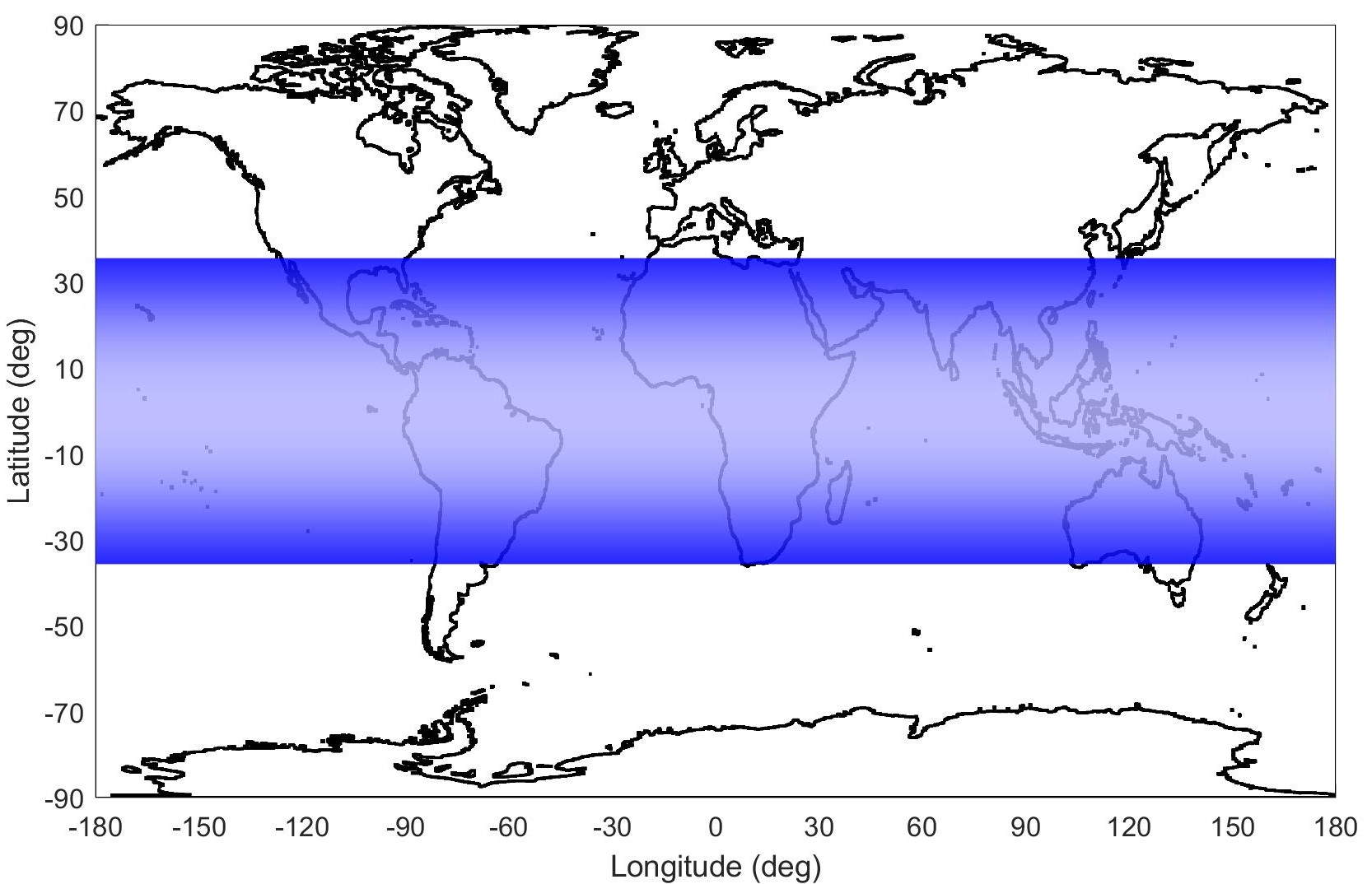}
\label{fig:GroundTrackDensity-b}
}
\refstepcounter{figure}\label{fig:GroundTrackDensity}
\end{figure}

\begin{figure}[htb]
\centering
\subfigure[The space tracks of an elliptical orbit. The tracks vary from yellow at perigee, to red at apogee. While it's more difficult to see the distribution of space tracks here than it is the for the ground tracks in \subref{fig:GroundTrackDensity-a}, they are very similar. We observe higher space track density at extreme latitudes just like before, with the main distinction being that we also observe a greater relative density of space tracks at perigee and apogee than in-between.]{
\centering
\includegraphics[width=.474\textwidth]{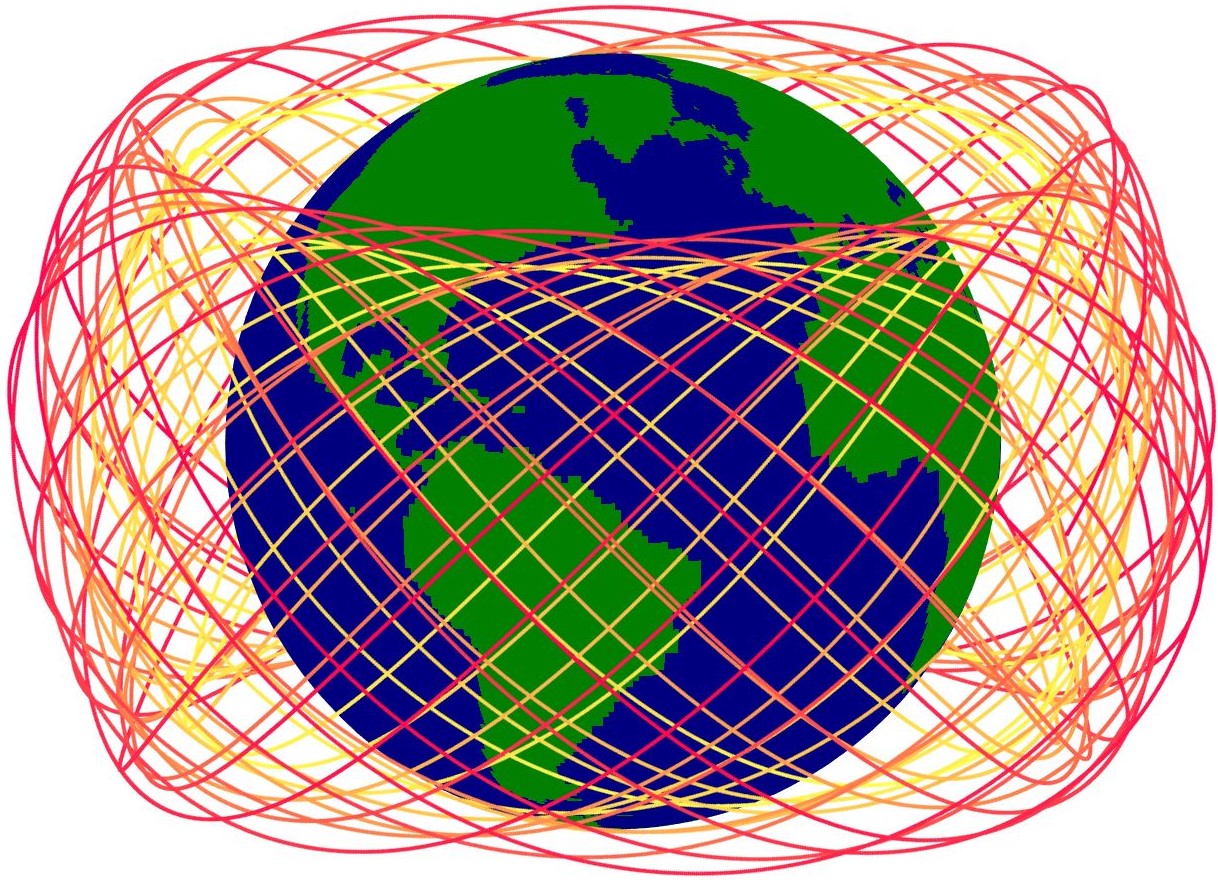}
\label{fig:SpaceTrackDensity-a}
}\hfill
\subfigure[The normalized ergodic probability density $\mu_{\text{e}}$, from Equation \ref{eqn:EllipticalMeasure}, of an elliptical orbit. Analogous to \subref{fig:GroundTrackDensity-b}, darker blue indicates a higher probability density. The analogy of the probability density in \subref{fig:GroundTrackDensity-b} to the density of space tracks in \subref{fig:SpaceTrackDensity-a} is less apparent on account of the higher-dimensionality and the variation in velocity with respect to altitude distorting the apparent distribution of space tracks.]{
\centering
\includegraphics[width=.464\textwidth]{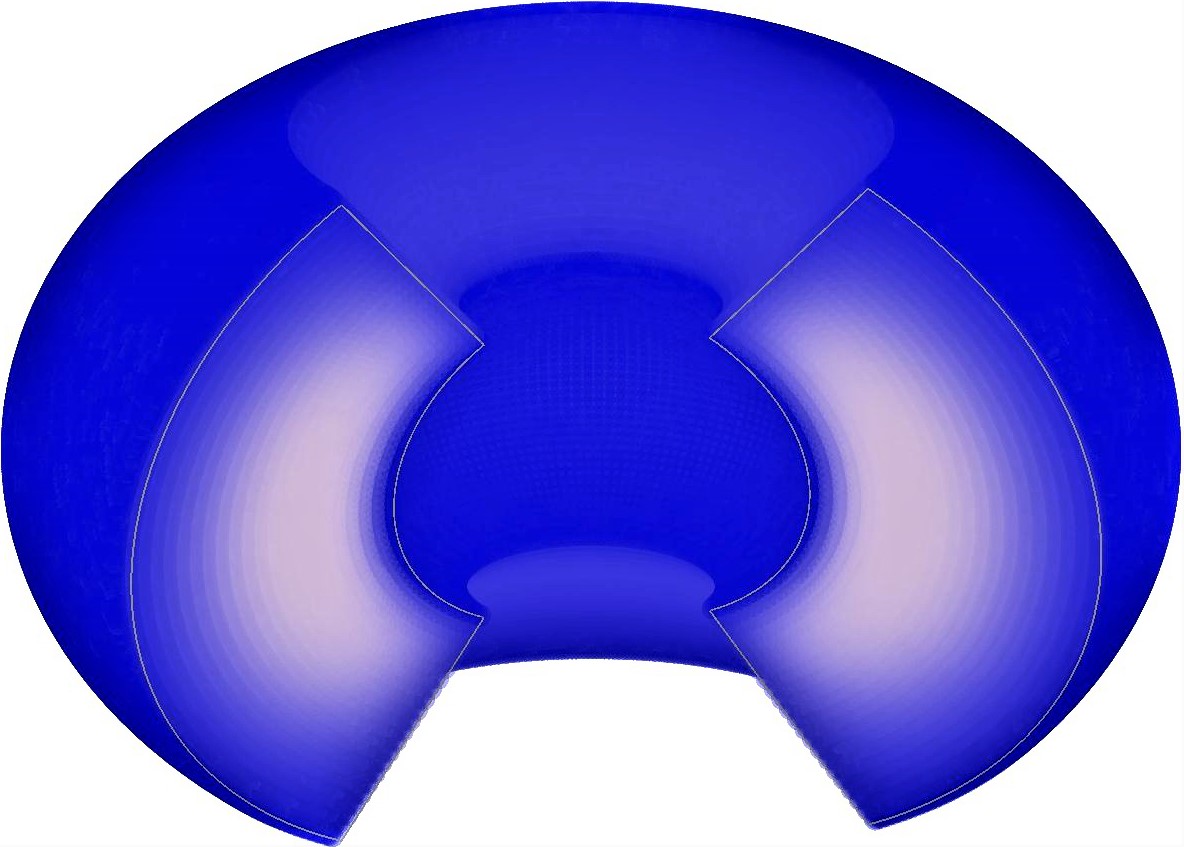}
\label{fig:SpaceTrackDensity-b}
}
\refstepcounter{figure}\label{fig:SpaceTrackDensity}
\end{figure}

Here, the dynamical system of interest is the $J_2-$perturbed 2 Body Problem, with $\phi_t(x)$ the position of the satellite over time. We can't expect $\mu$ to be uniform in this case, however. This is clear from the fact that the ground-tracks of satellite orbits are biased to extreme latitudes, as in \subref{fig:GroundTrackDensity-a}. In addition, for elliptical orbits one should expect a bias of the distribution towards larger radii due to the inverse relationship between satellite velocity and radial position \subref{fig:SpaceTrackDensity-a}. Equation \ref{eqn:EllipticalMeasure} from Graven and Lo 2019 \cite{GravenLo2019} provides the invariant measure for elliptical orbits. This extends the invariant measure for circular orbits from Lo 1994 \cite{Lo1994}, provided in Equation \ref{eqn:CircularMeasure}.\\\vspace{.1em}
\noindent\textbf{The Invariant Measure for Elliptical Satellite Orbits:}
\begin{align}
\begin{split}
\label{eqn:EllipticalMeasure}
\mu_{\text{e}}(r,\lambda,L&)=\dfrac{r\cos(\lambda )}{2\pi^3a\sqrt{{\sin^2(i)}-{\sin^2(\lambda )}}\sqrt{a^2e^2-(a-r)^2}}
\end{split}
\end{align}
\noindent\textbf{The Invariant Measure for Circular Satellite Orbits:}\vspace{-.2em}
\begin{align}
\begin{split}
\label{eqn:CircularMeasure}
\mu_{\text{c}}(\lambda,L&)=\dfrac{\cos(\lambda )}{2\pi^2\sqrt{\sin^2(i)-\sin^2(\lambda)}}
\end{split}
\end{align}

\footnotetext{Asymptotes on the boundary of the distribution remove most of the detail from the standard heat map, so we normalize the distribution to lie in $[0,1]$}
\vspace{-1em}
\section{Satellite-Ground Station View Period Ratios}
The original motivation for this work was accelerating the computation of the view period ratio, $\rho$. Lo 1994 \cite{Lo1994} provided this result for circular orbits. In the following sections, we demonstrate the utility of the result for elliptical orbits.
\FloatBarrier
\subsection{The General Problem Geometry}
\begin{figure}[htb]
\centering
\subfigure[The elevation angle constraint on the region of visibility. The elevation angle $\epsilon$ is the minimum communication angle above the horizon. The ground station mask angle, $\theta_0$, is the maximum communication angle between it and the satellite.]{
\centering
\includegraphics[width=.47\textwidth]{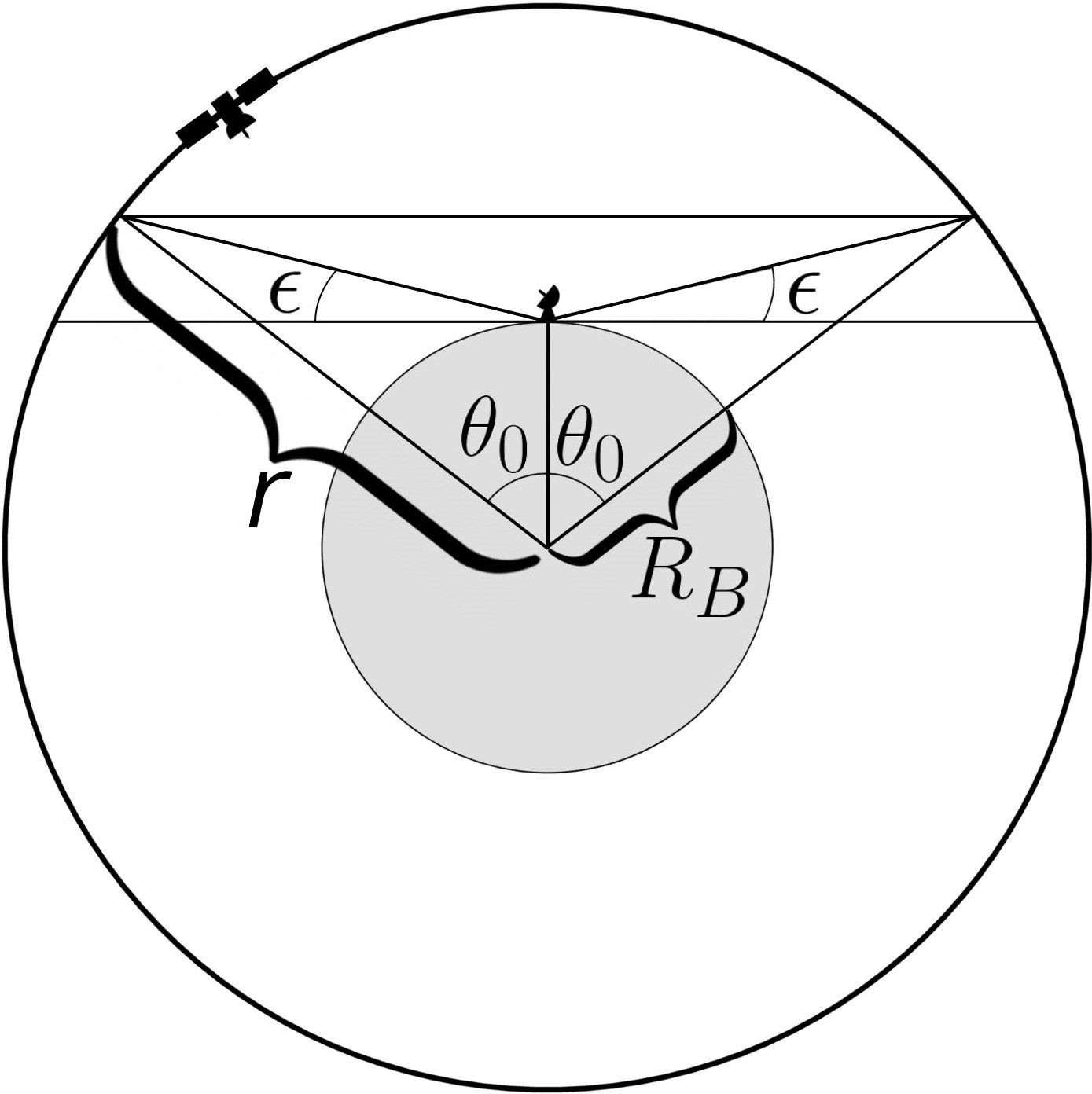}
\label{fig:GSMaskGeometry-a}
}\hspace{.4em}
\subfigure[Visualization of the satellite FOV constraint on the region of visibility from Equation \ref{eqn:Elev&FOVMaskAngle}. The FOV angle $\beta$ is the largest angle off-nadir at which the satellite can communicate. A formula is provided in Equation \ref{eqn:Elev&FOVMaskAngle}.]{
\centering
\includegraphics[width=.47\textwidth]{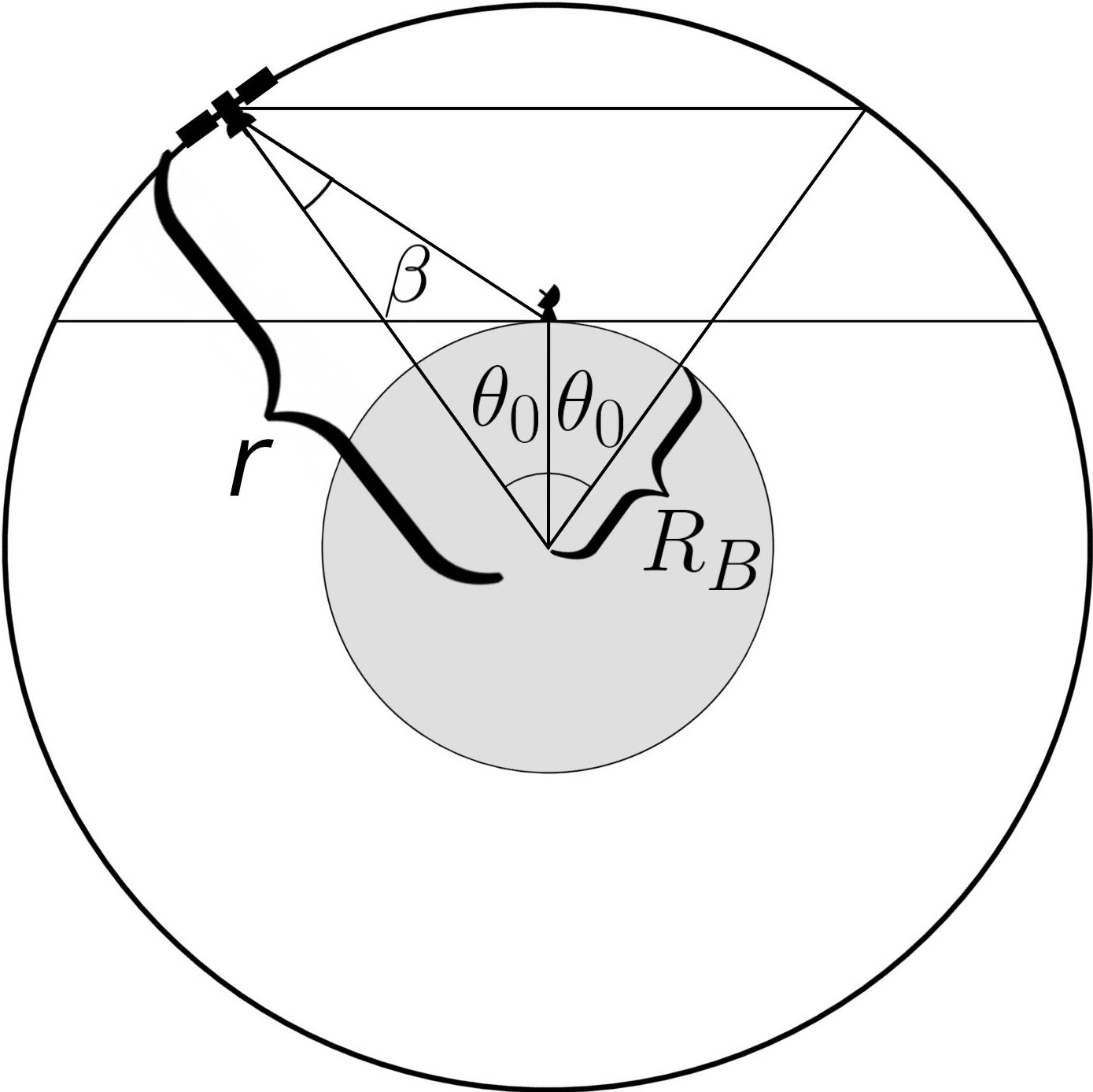}
\label{fig:GSMaskGeometry-b}
}
\refstepcounter{figure}\label{fig:GSMaskGeometry}
\end{figure}
Let $S$ be a satellite at some radius $r$, and suppose that its ground station has its line of sight constrained by an elevation angle $\epsilon$, such that it can only communicate with satellites at least $\epsilon$ radians above the horizon. Also suppose $S$ has a nadir-pointing communication instrument with an FOV angle constraint, $\beta$, such that $S$ can only send/receive within a cone of angular radius $\beta$ about nadir. Then, the communication geometry is as in Figure \ref{fig:GSMaskGeometry}. Applying standard trigonometric identities, we can describe each of these constraints by the parameter $\theta_{0}$, shown in Figure \ref{fig:GSMaskGeometry}.
\begin{align}
\begin{split}
\label{eqn:Elev&FOVMaskAngle}
\theta_{0,\text{elev}}(r)&=\cos^{-1}\left(\dfrac{R_B}{r}\cos(\epsilon)\right)-\epsilon\\
\theta_{0,\text{FOV}}(r)&=\begin{cases} 
    \sin^{-1}\left(\dfrac{r}{R_B}\sin(\beta)\right)-\beta, & \sin(\beta)<\dfrac{R_B}{r}\\
    \cos^{-1}\left(\dfrac{R_B}{r}\right), & \sin(\beta)\geq\dfrac{R_B}{r}\\
\end{cases}
\end{split}
\end{align}

And if both constraints are in effect, one can compute the effective ground station mask angle by taking the minimum of the two quantities from Equation \ref{eqn:Elev&FOVMaskAngle}.
\begin{align}
\begin{split}
\label{eqn:GSMaskAngle}
\theta_0(r)=\min\left\{\theta_{0,\text{elev}}(r),\theta_{0,\text{FOV}}(r)\right\}
\end{split}
\end{align}
Finally, given a ground station mask angular radius $\theta_0$, the ground station mask itself can be described in coordinates suitable for integration, as in Equation \ref{eqn:GSMask}.
\begin{align}
\begin{split}
\label{eqn:GSMask}
(\text{Ground Station Mask})(r)&=\big\{(r,\lambda,L)\big|\lambda\in\left[\lambda_{\min}(r),\lambda_{\max}(r)\right]\\
&\;\;\;\;\;\;\;L\in\left[-L_{\text{bound}}(\lambda,r),L_{\text{bound}}(\lambda,r)\right]\big\}\\
\lambda_{\min}(r)&=\trunc_i\left(g-\theta_0(r)\right)\\
\lambda_{\max}(r)&=\trunc_i\left(g+\theta_0(r)\right)\\
L_{\text{bound}}(\lambda,r)&=\cos^{-1}\left(\dfrac{\cos(\theta_0(r))-\sin(\lambda)\sin(g)}{\cos(\lambda)\cos(g)}\right)
\end{split}
\end{align}
Of course, the proportion of time the satellite spends in the ground station mask will be exactly the satellite-ground station view period ratio. Moreover, the invariant measure $\mu$ provides the infinitesimal proportion of time the satellite spends in each region of its state-space. Thus, the view period ratio can be computed as the integral over the ground station mask with respect to $\mu$. We will be referring to these formulae extensively in the following sections.

\subsection{One Satellite, One Ground Station}
First, we will consider the most basic case of a single satellite in a circular orbit communicating with a single ground station. This can then straightforwardly be extended to more complex geometries.
\subsubsection{The Circular Orbit View Period Formula}
Suppose we have a satellite in a circular orbit with semimajor axis $a$ and inclination $i$. Then, recalling equations \ref{eqn:CircularMeasure} and \ref{eqn:GSMask}, we can compute the satellite-ground station view period ratio, $\rho$ as:
$$\rho=\int\limits_{\substack{\text{Ground}\\\text{Station Mask}}}d\mu_{\text{c}}=\int\limits_{\lambda_{\min}}^{\lambda_{\max}}\int\limits_{-L_{\text{bound}}(\lambda)}^{L_{\text{bound}}(\lambda)}\mu_{\text{c}}(\lambda,L) dLd\lambda$$
Plugging in the formulae for $\lambda_{\min},\lambda_{\max},L_{\text{bound}}$ and $\mu_{\text{c}}$, and evaluating the inner integral yields:
$$\rho=\dfrac{1}{\pi^2}\int\limits_{\trunc_i(g-\theta_0)}^{\trunc_i(g+\theta_0)}\dfrac{\cos(\lambda)\cos^{-1}(\frac{\cos(\theta_0)-\sin(\lambda)\sin(g)}{\cos(\lambda)\cos(g)})}{\sqrt{\sin^2(i)-\sin^2(\lambda)}}d\lambda$$
Now, note that as the latitude approaches $\pm i$, the integrand blows up, and that singularities in the integrand tend to slow down numerical integration. To avoid this issue, we take the change of variables:
\begin{align}
\begin{split}
\label{eqn:LambdaChangeOfVariables}
\lambda(\alpha)=\sin^{-1}(\sin(i)\sin(\alpha))
\end{split}
\end{align}
This yields a formula for the satellite-ground station view period ratio for circular orbits in terms of a single definite integral, given in Equation \ref{eqn:CircularVPRFormula}. Accurate evaluation of this formula can be achieved with only 10 Gaussian quadrature nodes, and enables $\approx 287,000$ ratios/sec on a dual core 2.7GHz laptop CPU, or $\approx2.77\cdot10^{9}$ ratios/sec on the $5120$ core, $1.455GHz$ Titan V GPU.
\begin{figure*}[htb]
\begin{mdframed}
\begin{equation}
\begin{split}
\label{eqn:CircularVPRFormula}
\rho&=\dfrac{1}{\pi^2}\int\limits_{\alpha_1}^{\alpha_2}\cos^{-1}\left(\frac{\cos(\theta_0)-\sin(i)\sin(\alpha)\sin(g)}{\sqrt{1-\sin^2(i)\sin^2(\alpha)}\cos(g)}\right)d\alpha,\\
\alpha_1&=\sin^{-1}\left(\sin\left(\trunc_i\left(g-\theta_0\right)\right)\sin(i)^{-1}\right),\\
\alpha_2&=\sin^{-1}\left(\sin\left(\trunc_i\left(g+\theta_0\right)\right)\sin(i)^{-1}\right),\\
\theta_0&=\min\left\{\cos^{-1}\left(\dfrac{R_B}{a}\cos(\epsilon)\right)-\epsilon,\;\begin{cases} 
    \sin^{-1}\left(\dfrac{a}{R_B}\sin(\beta)\right)-\beta, & \sin(\beta)\leq\dfrac{R_B}{a}\\
    \cos^{-1}\left(\dfrac{R_B}{a}\right), & \sin(\beta)\geq\dfrac{R_B}{a}\\
\end{cases}\right\}
\end{split}
\end{equation}
\end{mdframed}
\captionsetup{labelformat=empty}
\vspace{-.6em}
\caption*{\textbf{Equation \ref{eqn:CircularVPRFormula}:} The simplified satellite-ground station view period ratio, $\rho$ formula for circular orbits. $a,i,\epsilon,\beta$ are the semimajor axis (km), orbital inclination (rad), ground station elevation angle (rad) and satellite FOV angle (rad) respectively.}
\vspace{-1em}
\end{figure*}
\newpage
\subsubsection{The Elliptical Orbit View Period Formula}
Now, suppose we have a satellite in an elliptical orbit with semimajor axis $a$, eccentricity $e$ and inclination $i$. Then, noticing that for each radius $r\in[a(1-e),a(1+e)]$, the cross-section of the cone of visibility is exactly described by Equation \ref{eqn:GSMask}. Thus, we can compute the view period ratio as the integral over these masks with respect to $\mu_{\text{e}}$:
$$\rho=\int\limits_{a(1-e)}^{a(1+e)}\left(\int\limits_{\lambda_1(r)}^{\lambda_2(r)}\int\limits_{-L(r,\lambda)}^{L(r,\lambda)}\mu_e(r,\lambda,L) dLd\lambda\right)dr$$
Applying the same approach as in the circular case, but extending the change of variables:
\begin{align}
\begin{split}
\label{eqn:ChangeOfVariables}
r(\theta)=a(1-e\sin(\theta)),\;\;\lambda(\alpha)=\sin^{-1}(\sin(i)\sin(\alpha))
\end{split}
\end{align}
The simplified view period ratio formula for the elliptical case is given in Equation \ref{eqn:EllipticalVPRFormula}.
\begin{figure*}[htb]
\begin{mdframed}
\begin{equation}
\begin{split}
\label{eqn:EllipticalVPRFormula}
\rho&=\dfrac{1}{\pi^3}\int\limits_{-\frac{\pi}{2}}^{\frac{\pi}{2}}\int\limits_{\alpha_1(\theta)}^{\alpha_2(\theta)}(1-e\sin(\theta))\cos^{-1}\left(\frac{\cos(\theta_0(\theta))-\sin(\alpha)\sin(i)\sin(g)}{\sqrt{1-\sin^2(\alpha)\sin^2(i)}\cos(g)}\right)d\alpha d\theta,\\
\alpha_1(\theta)&=\sin^{-1}\left(\sin\left(\trunc_i\left(g-\theta_0(\theta)\right)\right)\sin(i)^{-1}\right),\\
\alpha_2(\theta)&=\sin^{-1}\left(\sin\left(\trunc_i\left(g+\theta_0(\theta)\right)\right)\sin(i)^{-1}\right),\\
\theta_0(\theta)&=\min\left\{\cos^{-1}\left(P(\theta)\cos(\epsilon)\right)-\epsilon,\;\begin{cases} 
    \sin^{-1}\left(\dfrac{\sin(\beta)}{P(\theta)}\right)-\beta, & \sin(\beta)\leq P(\theta)\\
    \cos^{-1}\left(P(\theta)\right), & \sin(\beta)\geq P(\theta)\\
\end{cases}\right\}\\
P(\theta)&=\dfrac{R_B}{a(1-e\sin(\theta))}
\end{split}
\end{equation}
\end{mdframed}
\captionsetup{labelformat=empty}
\vspace{-.6em}
\caption*{\textbf{Equation \ref{eqn:EllipticalVPRFormula}:} The simplified satellite-ground station view period ratio formula for elliptical orbits. $a,e,i,\epsilon,\beta$ are the semimajor axis (km), eccentricity (unitless), orbital inclination (rad), ground station elevation angle (rad) and satellite FOV angle (rad) respectively.}
\end{figure*}
\subsection{One Satellite, Many Ground Stations}
The introduction of additional, potentially overlapping, ground stations turns out to non-additively increase the complexity of computing the view period ratio. We conceptualize this problem as follows: Given a set of $N$ ground stations and a satellite trajectory, we can represent each ground station by a latitude, longitude, ground station mask angle triple:
$$(\lambda_1,L_1,\theta_1),(\lambda_2,L_2,\theta_2),\hdots,(\lambda_N,L_N,\theta_N)$$
and the satellite trajectory by its orbital elements $(a,e,i)$. Using this information we'd like to determine the expected total visibility time for the satellite with the complete set of ground stations. We will begin with the circular case.
\subsubsection{The Circular Case}
\begin{figure}[htb]
\centering
\includegraphics[width=.8\textwidth]{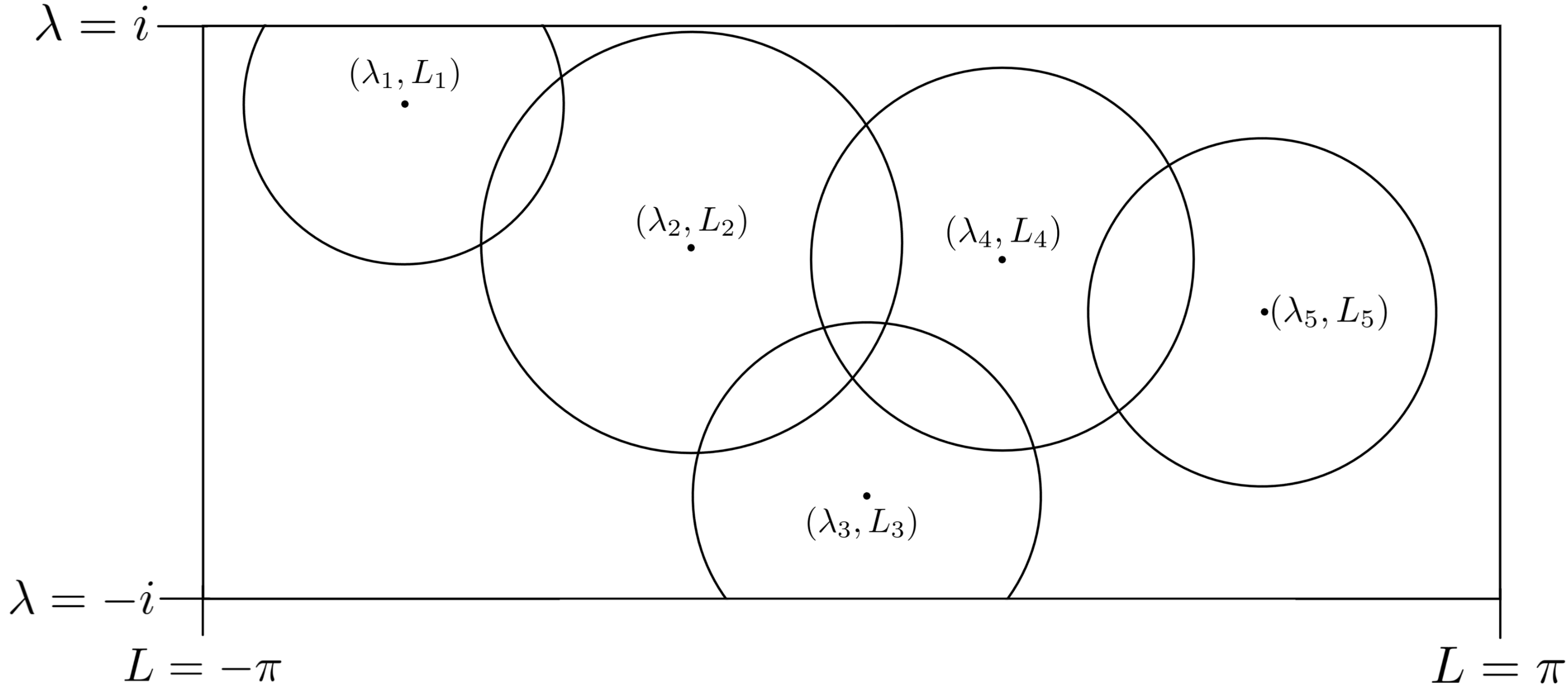}
\caption{Example set of ground station masks for multiple ground stations}
\label{fig:MultipleGSMasks}
\end{figure}

Here, we are given the information outlined above, except the satellite trajectory is circular, so we only need its semimajor axis $a$ and inclination $i$. For example, the situation for $N=5$ may appear as in Figure \ref{fig:MultipleGSMasks}.
\\
\forceindent Note that we can't directly compute the view period ratio for each ground station and sum them because the ground station masks may overlap, resulting in the double counting of some regions. With a little work, it is possible to take this into account. The view period ratio formula in this case is given by Equation \ref{eqn:CircularMultipleGSVPR}.
\begin{align}
\begin{split}
\label{eqn:CircularMultipleGSVPR}
\rho&=\dfrac{1}{2\pi^2}\int\limits_{\alpha_{\min}}^{\alpha_{\max}}\text{Length}(I(\lambda(\alpha)))d\alpha\\
\alpha_{\min}&=\sin^{-1}\left(\dfrac{\sin\trunc_i\left(\min\limits_k\{\lambda_k-\theta_k\}\right)}{\sin(i)}\right)\\
\alpha_{\max}&=\sin^{-1}\left(\dfrac{\sin\trunc_i\left(\min\limits_k\{\lambda_k+\theta_k\}\right)}{\sin(i)}\right)
\end{split}
\end{align}
Dealing with the complexities introduced by the overlapping ground station masks is a lengthy process. Thus, the explanation of Equation \ref{eqn:CircularMultipleGSVPR}, including the definition of definition of $\text{Length}(I(\lambda(\alpha)))$ has been relegated to Appendix A. Importantly, computing $\text{Length}(I(\lambda(\alpha)))$ is at worst $O(N\log(N))$ if $N$ is the number of ground stations. Thus, we only pick up a logarithmic term in the computational complexity, in comparison to the case of non-overlapping ground stations.

\subsubsection{The Elliptical Case}
Taking a similar approach, the formula in the elliptical case is given by:
\begin{align}
\begin{split}
\label{eqn:EllipticalMultipleGSVPR}
\rho&=\dfrac{1}{2\pi^3}\int\limits_{-\frac{\pi}{2}}^{\frac{\pi}{2}}\int\limits_{\alpha_{\min}(r)}^{\alpha_{\max}(r)}(1-e\sin(\theta))\text{Length}(I(\lambda(\alpha),r(\theta)))d\alpha d\theta\\
\alpha_{\min}(r)&=\sin^{-1}\left(\dfrac{\sin\trunc_i\left(\min\limits_k\{\lambda_k-\theta_k(r)\}\right)}{\sin(i)}\right)\\
\alpha_{\max}(r)&=\sin^{-1}\left(\dfrac{\sin\trunc_i\left(\min\limits_k\{\lambda_k+\theta_k(r)\}\right)}{\sin(i)}\right)
\end{split}
\end{align}

\subsection{Many Satellites, One (or Many) Ground Station(s)}
The natural next step after considering the single satellite-single ground station and single satellite-multiple ground station cases is considering what can be said of the coverage if we're working with an ensemble satellites and one (or several) ground station(s). If the satellite orbits are statistically dependent, then this appears to be a difficult problem. However under the assumption of independence, a lot can be said using only the view period formulae from the previous sections.
\subsubsection{Two Satellites, One Ground Station}
Suppose we have two satellites, $S_1$ and $S_2$, and a ground station, $G$. Also suppose that the view period ratios for $S_1$ and $S_2$ are $\rho_1$ and $\rho_2$ respectively. Then, note that we can view $\rho_1,\rho_2\in[0,1]$ as the instantaneous probability that either of the satellites will be visible to the ground station at any point in time. Then, because their trajectories are independent: 
\begin{align}
\begin{split}
\label{eqn:DblVisOverlap}
P(S_1\text{ and }S_2\text{ visible to }G)&=P(S_1\text{ visible to }G)P(S_2\text{ visible to }G)=\rho_1\rho_2
\end{split}
\end{align}
Similarly, the probability that exactly one of the satellites is in view of the ground station are $\rho_1(1-\rho_2)$ and $\rho_2(1-\rho_1)$ respectively. Moreover, we can compute the total coverage ratio:
\begin{align}
\begin{split}
\label{eqn:DblVis}
P(S_1\text{ or }S_2\text{ visible to }G)&=\rho_1+\rho_2-\rho_1\rho_2
\end{split}
\end{align}
These extended coverage ratios have a very natural interpretation: They represent the expected proportion of flight time for which the given condition holds. (e.g. $\rho_1\rho_1$ is the expected proportion of flight time for which we can expect both satellites to have line-of-sight with $G$. Similarly, $\rho_1+\rho_2-\rho_1\rho_2$ is the expected proportion of flight time for which we can expect at least one satellite to have line-of-sight with $G$.)
\subsubsection{$N$ Satellites, One Ground Station}
From here, it's straightforward to generalize to $N$ satellites $S_1,S_2,\hdots,S_N$ in communication with a single ground station, $G$, each satellite with a view period ratio, $\rho_i$. Firstly, we can extend Equation \ref{eqn:DblVisOverlap} to compute the probability that any subset $S_{\{i_1,\hdots,i_{N'}\}}=\{S_{i_1},S_{i_2},\hdots,S_{i_{N'}}\}$ of the satellites is visible to $G$:
\begin{align}
\begin{split}
\label{eqn:NVisOverlap}
P\left(S_{\{i_1,\hdots,i_{N'}\}}\text{ visible to }G\right)&=P\left(\bigcap\limits_{i\in\{i_1,\hdots,i_{N'}\}}S_i\text{ visible to }G\right)=\prod\limits_{i\in\{i_1,\hdots,i_{N'}\}}\rho_i
\end{split}
\end{align}
Next, using $P(A)=1-P(\neg A)$, we can see the overall coverage ratio of the $N$ satellites:
\begin{align}
\begin{split}
\label{eqn:NVis}
P\left(\geq1\text{ of the }S_i\text{ visible to }G\right)&=1-\prod_{i=1}^N(1-\rho_i)
\end{split}
\end{align}
Clearly these formulae can be applied analogously when there are multiple ground stations simply by replacing the single ground station view period ratios with the multiple ground station view period ratios. In addition, in the multiple ground station case, these formulae can be applied to analyze individual ground stations and, in general, any subset of the ground stations under consideration.
\section{The Long-Term Mean Value of Functions of Satellite Position}
Recall that the Birkhoff-Kinchin Theorem asserts a time mean-space mean equivalence for ergodic dynamical systems. This relation is given explicitly in Equation \ref{eqn:BirkhoffKinchin}. In the case of computing view period ratios, the formulae provided in the previous sections can be thought of as applications of the theorem to $f=\mathbf{1}_{\text{Vis}}$ (the indicator function on the region of visibility of the ground station(s)), and $\vec{x}(t)$ the position of the satellite as a function of time. But this formula actually applies in much greater generality. In particular, $f:S\rightarrow\mathbb{R}^n$ can be any measurable scalar or vector-valued (and even possibly tensor-valued) function on the state space of the satellite. For example:
\begin{itemize}[topsep=0pt]
    \item Satellite speed: $v:S\rightarrow\mathbb{R}$, ie. $v(r,\lambda,L)=\sqrt{\mu\left(\dfrac{2}{r}-\dfrac{1}{a}\right)}$.
    \item Atmospheric density: $\rho:S\rightarrow\mathbb{R}$.
    \item Drag force per unit area: $D:S\rightarrow\mathbb{R}$, ie. $D(r,\lambda,L)=Cd\dfrac{\rho(r,\lambda,L)v(r,\lambda,L)^2}{2}$.
    \item The magnetic field strength or direction: $|B|:S\rightarrow\mathbb{R}$, $\;\;\;B:S\rightarrow\mathbb{R}^3$.
    \item The gravity gradient tensor: $\Gamma:S\rightarrow\mathbb{R}^{3\times3}$.
\end{itemize}
It's important to note that the standard form of the Birkhoff-Kinchin Theorem assumes that the function being averaged can be made to depend exclusively on the spatial state of the system. However, many functions of interest are not necessarily time independent. For example, both the Earth's atmospheric density and magnetic field fluctuate over time. In cases such as this, mean values for each location in $S$ can be considered as an approximation - although this will be application dependent and requires further validation.
\subsection{Circular Orbits}
\noindent Applying the Birkhoff-Kinchin theorem, we can straightforwardly write down an integral formula for the long term time mean of any measurable function $f$ of the position of the satellite:
\begin{align}
\begin{split}
\label{eqn:MeanFCirc}
E_{\mu_{\text{c}}}(f)=\int\limits_{-i}^i\int\limits_{-\pi}^{\pi}\mu_{\text{c}}(\lambda,L)f(r,\lambda,L)dLd\lambda
\end{split}
\end{align}
It then becomes a triviality to compute higher order quantities such as variance:
\begin{align}
\begin{split}
\label{eqn:VarFCirc}
\text{Var}_{\mu_{\text{c}}}(f)=\int\limits_{-i}^i\int\limits_{-\pi}^{\pi}\mu_{\text{c}}(\lambda,L)\left(f(r,\lambda,L)-E_{\mu_{\text{c}}}(f)\right)^2dLd\lambda
\end{split}
\end{align}
(With the square taken entry-wise if $f$ isn't scalar-valued)\\
Also note that because the probability measure has no dependence on the longitude, if $f$ does as well, then we can directly evaluate the inner-most integral in each case, reducing these to single integrals. A similar approach can be used for latitude in some cases as well.

\subsection{Elliptical Orbits}
\noindent Doing the same as in the previous section, but for the elliptical measure:
\begin{align}
\begin{split}
\label{eqn:MeanF}
E_{\mu_e}(f)=\int\limits_{a(1-e)}^{a(1+e)}\int\limits_{-i}^i\int\limits_{-\pi}^{\pi}\mu(r,\lambda,L)f(r,\lambda,L)dLd\lambda dr
\end{split}
\end{align}
We can compute the variance in this case as well:
\begin{align}
\begin{split}
\label{eqn:VarF}
\text{Var}_{\mu_e}(f)=\int\limits_{a(1-e)}^{a(1+e)}\int\limits_{-i}^i\int\limits_{-\pi}^{\pi}\mu(r,\lambda,L)\left(f(r,\lambda,L)-E_\mu(f)\right)^2dLd\lambda dr
\end{split}
\end{align}
Just as in the previous section, it will often be possible to reduce these volume integrals to single or double integrals by directly evaluating one or more of the integrals in the expression. There can be significant benefit to doing so, because accurate evaluation of volume integrals can be a numerically costly operation.

In many cases, it's possible to evaluate these formulae in closed form. For example, we can apply Equation \ref{eqn:MeanF} to easily compute the mean radius of of an elliptical orbit with respect to time:
$$E_\mu(r)=\int\limits_{a(1-e)}^{a(1+e)}\int\limits_{-i}^i\int\limits_{-\pi}^{\pi}\mu(r,\lambda,L)rdLd\lambda dr=\dfrac{a}{\pi}\int\limits_{\theta=-\frac{\pi}{2}}^{\frac{\pi}{2}}(1-e\sin(\theta))^2d\theta=a\left(\dfrac{e^2}{2}+1\right)$$
Which is in agreement with the standard formula.\cite{MeanRad}

\section{Numerical Results}
In this section, numerical accuracy results for some of the preceding formulae are provided. In general, the ``ground truth'' we compare these numerical results against is the value calculated via direct trajectory propagation on an RK-78 integrator, using the $J_2$-perturbed force model provided in Equation \ref{eqn:NonlinMotionEqs}. The absolute error and percent error metrics used here are defined in equation \ref{eqn:ErrorMetrics}.
\begin{align}
\begin{split}
\label{eqn:ErrorMetrics}
\text{Absolute Error := }&\left|\text{True Value}-\text{Estimate}\right|\\
\text{Percent Error := }&\dfrac{\left|\text{True Value}-\text{Estimate}\right|}{\text{True Value}}\cdot100=\dfrac{\text{Absolute Error}}{\text{True Value}}\cdot100
\end{split}
\end{align}
Both of these metrics are utilized throughout this section, but primarily absolute error. To illustrate the reason for this, we consider the ``One Satellite, One Ground Station'' view period ratio case. Here, note that each view period ratio will be a number $\rho\in[0,.5]$. If $\rho=.25$, for example, this would imply that the satellite and ground station can communicate approximately $25\%$ of the time. Now, suppose that the estimate given by Equation \ref{eqn:EllipticalVPRFormula} were $.24$. Then, the absolute error $=|.25-.24|=.01$ has a very natural interpretation: the view period ratio estimate from the formula is off by $1\%$ of total flight time. Similarly, the percent error $=\frac{|.25-.24|}{.25}\cdot100=4\%$ has the interpretation: the view period ratio estimate from the formula deviates by $4\%$ from the true view period ratio. In this case, each of these work as effective metrics. However, consider a slightly different case, where $\rho_{\text{true}}=.005$, and $\rho_{\text{estimate}}=.01$. Then, the absolute error $=|.005-.01|=.005$, and the percent error $=\frac{.005}{.005}\cdot100=100\%$. In this scenario, we can still interpret these values in the same way as before, however while the true and estimate values are far closer than in the first case, the percent error metric represents $\rho_{\text{estimate}}$ as a very poor estimate. This issue only worsens as $\rho_{\text{true}}\to0$, and if $\rho_{\text{true}}=0$, then the percent error isn't well-defined. Thus, although there may be cases where percent error is a useful metric, absolute error is preferred here due to its robustness.

\subsection{One Satellite, One Ground Station}
The numerical results given below demonstrate the distribution of the absolute error of the view period ratio formulae given by equations \ref{eqn:CircularVPRFormula} and \ref{eqn:EllipticalVPRFormula} when compared against the method of direct orbit propagation for 6000 (integration) days, using the $J_2$-perturbed Earth model. Each of the 50,000 cases given were randomly sampled from the following sample space:\vspace{-.3em}
\begin{align*}
    &\text{Semimajor Axis} &a&\in[6371.0088+160,20000]\text{ (km)}\\
    &\text{Eccentricity} &e&\in[0,.6]\text{ (unitless)}\\
    &\text{Orbital Inclination} &i&\in[0,90]\text{ (deg)}\\
    &\text{Ground Station Latitude} &g&\in[0,90]\text{ (deg)}\\
    &\text{Elevation Angle} &\epsilon&\in[0,50]\text{ (deg)}\\
    &\text{Field of View Angle} &\beta&\in[0,90]\text{ (deg)}
\end{align*}
In addition, we require that the apogee is at least 160km above the surface.\vspace{-.5em}

\begin{figure}[htb]
\centering
\subfigure[A histogram of the absolute error, defined in Equation \ref{eqn:ErrorMetrics}, for the 50,000 cases assigned above, with vertical lines denoting the values of absolute error such that $50\%,85\%,95\%$ and $99\%$ of the cases have error less than that value. In addition, cases with error $\geq.1$ account for $.366\%$ of the total of cases.]{
\centering
\includegraphics[width=.47\textwidth]{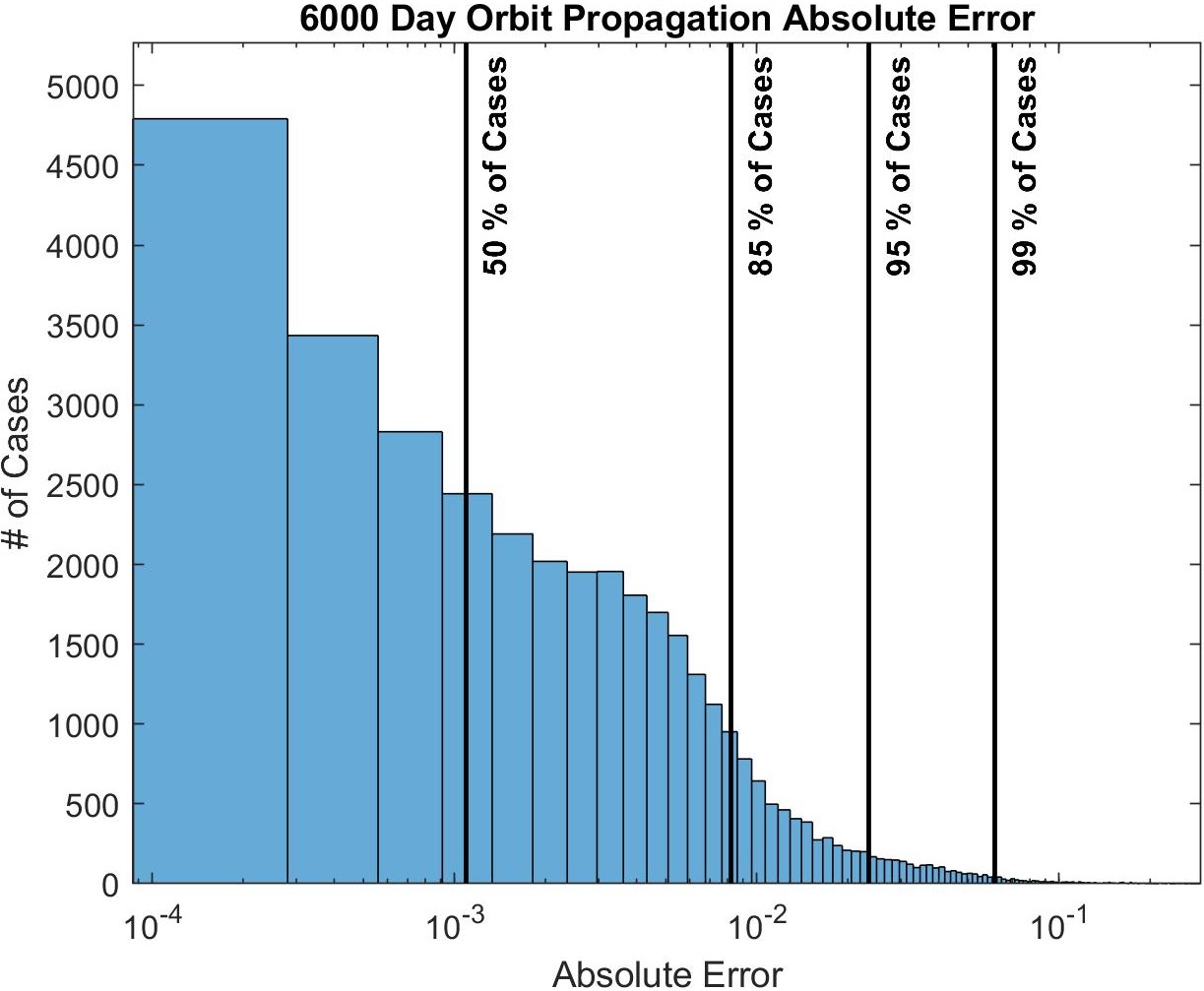}
\label{fig:AbsErrHist1}
}\hfill
\subfigure[The same as \subref{fig:AbsErrHist1}, the difference being that the cases with orbital inclination within $1.5^\circ$ are filtered out (amounting to 3.408\% of the $50,000$ cases). Notice that with these critically (or nearly critically) inclined orbits filtered out, each of the percentiles of cases occurs earlier. These differences are most pronounced for the $95\%$ and $99\%$ thresholds.]{
\includegraphics[width=.48\textwidth]{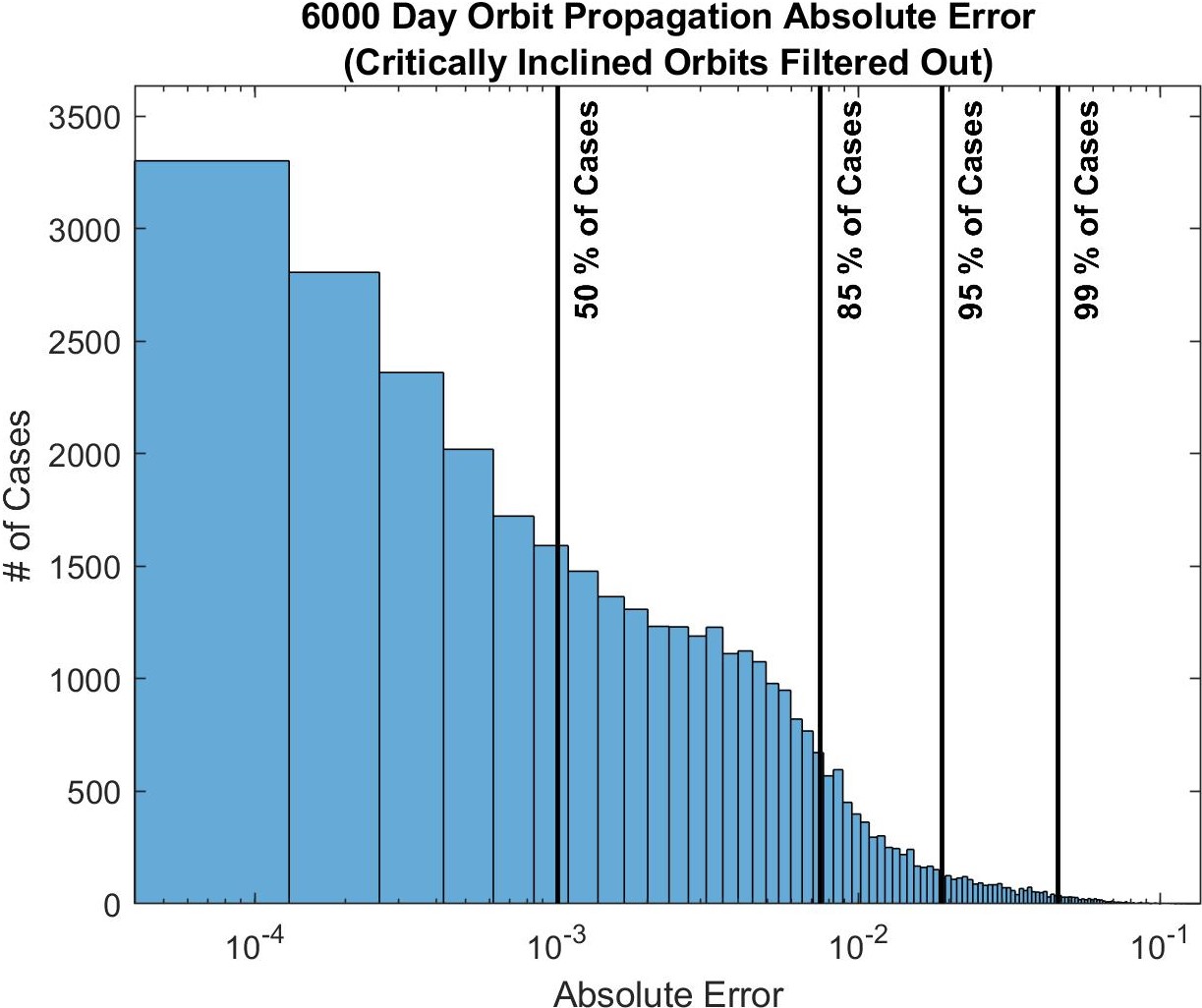}
\label{fig:AbsErrHistNoCrit}
}
\refstepcounter{figure}\label{fig:AbsErrHist}
\end{figure}

\begin{figure}[htb]
\centering
\includegraphics[width=.80\textwidth]{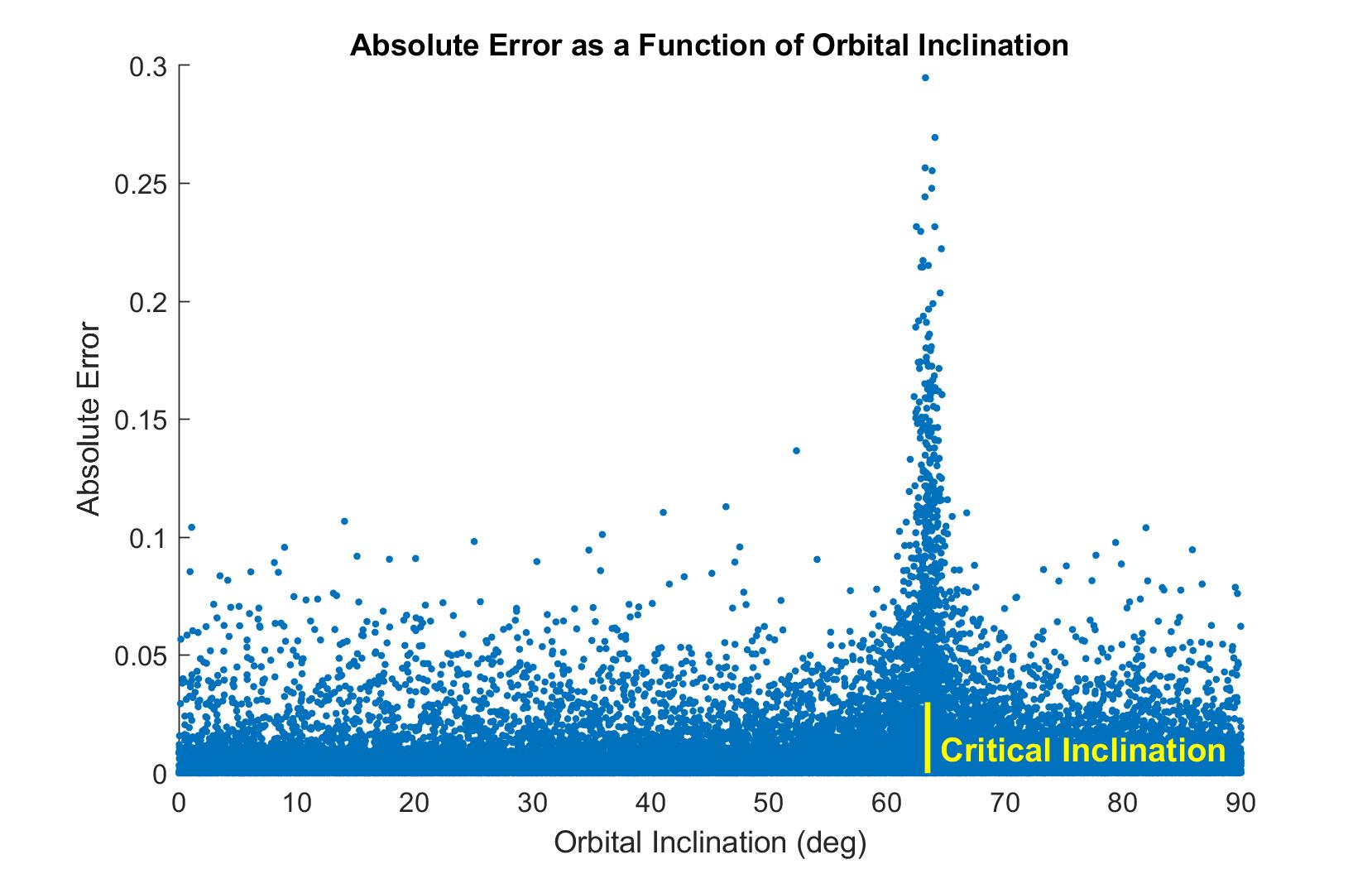}
\caption{A scatter plot of the absolute error, defined in Equation \ref{eqn:ErrorMetrics}, for each of the 50,000 cases as a function of the inclination of the orbit. Note the significant increase in error at critical inclination, which occurs at $\text{inclination}\approx63.43^\circ$. Critical inclination is the orbit inclination at which the satellite orbit experiences zero apogee drift, which constitutes a degenerate case for the view period formula, so this was to be expected. Figure \ref{fig:AbsErrHist} illustrates the effect removing these cases has on the error profile.}
\label{fig:AbsErrCritInc}
\end{figure}

\begin{figure}[htb]
\centering
\includegraphics[width=.7\textwidth]{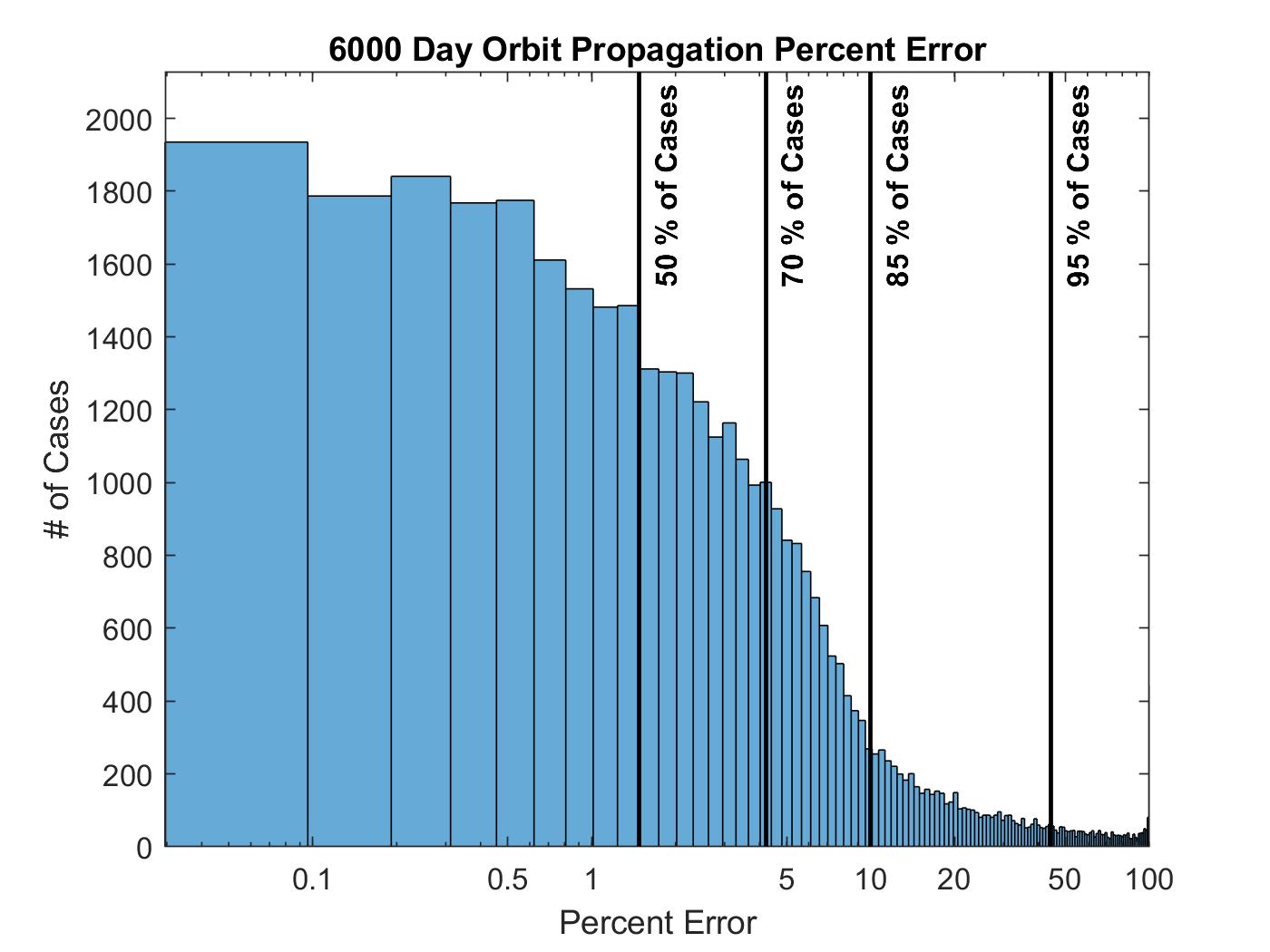}
\caption{Similar to the plots in Figure \ref{fig:AbsErrHist}, showing percent error, defined in Equation \ref{eqn:ErrorMetrics}. Cases where $\rho_{\text{true}}=0$, but $\rho_{\text{estimate}}\neq0$ were handled by setting them to $100\%$ error. This occurred in $31$ of the $50,000$ cases (or .68\%).}
\label{fig:PctErrHist1}
\end{figure}
\FloatBarrier
\section{Conclusion \& Future Work}
In this work, we provided an analytical method of estimating the long-term mean value of any function of satellite position, under the assumption of a $J_2-$perturbed and aperiodic orbit. Special emphasis was placed on applying this approach to rapid ground coverage assessment, with integral formulae optimized for numerical computation provided in Equations \ref{eqn:CircularVPRFormula}, \ref{eqn:EllipticalVPRFormula}, \ref{eqn:CircularMultipleGSVPR} and \ref{eqn:EllipticalMultipleGSVPR}. In particular, formulae for computing coverage while taking into account ground station elevation angle, satellite field of view, and multiple ground stations for both circular and elliptical satellite orbits were provided. Note that these formulae are really just a special case of the general method given in Equations \ref{eqn:MeanFCirc}, \ref{eqn:MeanF}, \ref{eqn:VarFCirc} and \ref{eqn:VarF} for estimating the time mean and variance of any function of the satellite trajectory. We anticipate that this will enable accelerated evaluation of many relevant flight parameters such as velocity, drag, magnetic field strength/direction, gravity gradient and sun $\beta-$angle, among other quantities of interest. It's worth noting that each of these formulae are completely independent of any physical constants, other than the radius of the body $R_B$, and (implicitly) $J_2$, which is necessary to determine the periodicity (or the lack thereof) of a given orbit.

The feasibility of extending this approach to quantifying the dynamics of multiple satellites was also investigated. The main barrier to this extension was the requirement that the satellite trajectories be statistically independent - a property which is unlikely to hold for most satellite constellations and formations, in which the orbits are often commensurate by design. The authors are actively working on extending the underlying theory to more effectively handle these cases. It's anticipated that this can be accomplished without significant increases in complexity or computational cost.

As was alluded to in the ``Many Satellites, One (or Many) Ground Station(s)'' section, it can be instructive to view, $\mu_c$ and $\mu_e$ as probability densities on the instantaneous position of the satellite. On this interpretation, it becomes natural to consider quantities such as expectation and variance. In addition, it suggests applications to orbit determination via maximum likelihood estimation (e.g. for situational awareness or exoplanet TLE estimation) and the utilization of Chebyshev-like bounds on the distribution of values of functions of satellite position. However further research will be necessary to accurately assess the efficacy of these and the limitations of the probabilistic interpretation of $\mu_c$ and $\mu_e$.

Throughout this paper, a linearized $J_2$ model was assumed. The $J_2-$perturbed two body problem may be well-approximated by a linear precession of the orbital elements Mean Anomaly, Argument of Periapsis, and Longitude of the Ascending Node. This linearization is a key step in the derivation of the results provided here. However, because this linearization only approximately captures the behavior of the system, with bias towards the secular effects, there are classes of orbits for which our method performs rather poorly. The most notable of which is the class of critically inclined orbits. As we saw in the numerical results section, there's a sharp increase in view period ratio estimation error for orbits which are near critical inclination ($\approx63.43^\circ$). Critically inclined orbits often exhibit pathological behavior in their own right\cite{critInc}, but our model introduces additional inaccuracies. In particular, in order for the ergodicity assumption on an orbit to be satisfied, it must be aperiodic which, in this context, means that the three linear precession rates ($\dot{M},\dot{\omega},\dot{\Omega}$) are rationally independent\footnote{Specifically, $\nexists \alpha\in\mathbb{Q}$ such that $\frac{\dot{M}}{\dot{\omega}}=\alpha$ or $\frac{\dot{\omega}}{\dot{\Omega}}=\alpha$ or $\frac{\dot{\Omega}}{\dot{M}}=\alpha$}. While such orbits constitute a set of measure zero, near-periodic orbits tend to exhibit reduced accuracy. Critical inclination is a special case of periodicity in which $\dot{\Omega}=0$, resulting in particularly poor accuracy in its vicinity. The use of a higher order (non-linearized) model would likely ameliorate many of these accuracy issues by capturing the effect of critical inclination.

\section{Acknowledgments}
This research was carried out in part at the Jet Propulsion Laboratory, California Institute of Technology under a contract with the National Aeronautics and Space Administration (80NM0018D0004). This work was sponsored in part by the Caltech Summer Undergraduate Research Fellowship Program. This work was also supported in part by the Hummer-Tuttle gift to Professor Al Barr through the Caltech Division of Engineering and Applied Science.
\section{Notation}
{\renewcommand\arraystretch{1.0}
\noindent\begin{longtable}{@{}l @{\quad=\quad} l@{}}
    $\rho$ & The View Period Ratio, $\rho\in[0,1)$\\
    $\mu$ & The Standard Gravitational Parameter\\
    $\mu_c$ & The Invariant Measure for Circular Orbits\\
    $\mu_e$ & The Invariant Measure for Elliptical Orbits\\
    $a$ & Orbit Semimajor Axis\\
    $e$ & Orbit Eccentricity\\
    $i$ & Orbit Inclination Angle\\
    $M$ & Mean Anomaly\\
    $\Omega$ & Longitude of the Ascending Node\\
    $\omega$ & Argument of Periapsis\\
    $r$ & The Radial Component of Satellite Position, $r\in(a(1-e),a(1+e))$\\
    $\lambda$ & The Latitude Component of Position, $\lambda\in[-i,i]$\\
    $L$ & The Longitude Component of Position $L\in[-\pi,\pi]$ rad or $L\in[-180,180]^\circ$\\
    $R_B$ & The Radius of the Central Body\\
    $J_2$ & The Second Zonal Harmonic\\
    $g$ & The Latitude of the Ground Station $g\in(-\frac{\pi}{2},\frac{\pi}{2})$\\
    $\epsilon$ & The Elevation Angle of the Ground Station, $\epsilon\in(0,\frac{\pi}{2})$\\
    $\beta$ & The Field of View (FOV) Angle of the Satellite, $\beta\in(0,\frac{\pi}{2})$\\
    $\theta_0$ & The Ground Station Mask Radius\\
    $S$ & The State Space of the Satellite, $S=[a(1-e),a(1+e)]\times[-i,i]\times[-\pi,\pi]$\\
    $\Omega_B$ & The Precession Rate of the Central Body\\
    $\text{E}_{\mu_{c}}(f)$ & The Expected Value of $f$ With Respect to the Probability Measure $\mu_c$ For Circular Orbits\\
    $\text{E}_{\mu_{e}}(f)$ & The Expected Value of $f$ With Respect to the Probability Measure $\mu_e$ For Elliptical Orbits\\
    $\text{Var}_{\mu_{c}}(f)$ & The Variance of $f$ With Respect to the Probability Measure $\mu_c$ For Circular Orbits\\
    $\text{Var}_{\mu_{e}}(f)$ & The Variance of $f$ With Respect to the Probability Measure $\mu_e$ For Elliptical Orbits\\
    $\trunc_b(x)$ & The Truncation Function, $\trunc_b(x)=\min\{\max\{x,-b\},b\}$
\end{longtable}}
\appendix
\section*{Appendix A}
Note that we can't just compute the naive view period ratio for each ground station and sum them because the ground station masks may overlap, as in the Figure \ref{fig:MultipleGSMasks}, resulting in double counting some regions.  We will need to do a bit of additional work to avoid that issue. Firstly, note that the naive view period integral for ground station $k$ with ground station mask radius $\theta_k$ is given by:
$$\rho=\int_{\lambda=\trunc_i(\lambda_k-\theta_k)}^{\trunc_i(\lambda_k+\theta_k)}\int_{L=L_k-\cos^{-1}(\frac{\cos(\theta_k)-\sin(\lambda)\sin(\lambda_k)}{\cos(\lambda)\cos(\lambda_k)})}^{L_k+\cos^{-1}(\frac{\cos(\theta_k)-\sin(\lambda)\sin(\lambda_k)}{\cos(\lambda)\cos(\lambda_k)})}\dfrac{\cos(\lambda)}{2\pi^2\sqrt{\sin^2(i)-\sin^2(\lambda)}}dLd\lambda$$
In particular for each value of $\lambda$ in range, we want to integrate over the values of $L$ in the interval:
$$[L_{k,\min}(\lambda),L_{k,\max}(\lambda)]:=$$
$$\left[L_k-\cos^{-1}\left(\dfrac{\cos(\theta_k)-\sin(\lambda)\sin(\lambda_k)}{\cos(\lambda)\cos(\lambda_k)}\right),L_k+\cos^{-1}\left(\dfrac{\cos(\theta_k)-\sin(\lambda)\sin(\lambda_k)}{\cos(\lambda)\cos(\lambda_k)}\right)\right]$$
Thus, when there are multiple ground stations, we want to integrate over the union of the intervals:
$$L_{\text{range}}(\lambda)=\bigcup_{k=1}^N[L_{k,\min}(\lambda),L_{k,\max}(\lambda)]$$
The simplest way to work with this union of intervals programmatically is to combine them into a union of disjoint intervals. There is a simple algorithm for doing so.\\\\
\textbf{Algorithm: Interval Merge\cite{IntervalMerge}}\\
Given a set of $N$ intervals: $[x_1,y_1],[x_2,y_2],\hdots,[x_N,y_N]$:
\begin{enumerate}
    \item Reorder the intervals in increasing order based on the lower bounds of the intervals. That is:
    $$[x_1,y_1],[x_2,y_2],\hdots,[x_N,y_N]\rightarrow [x_1',y_1'],[x_2',y_2'],\hdots,[x_N',y_N'],\text{ s.t: }x_1'\leq x_2'\leq\hdots\leq x_N'$$
    \item Push the first interval onto the stack.
    \item For each interval in the ordered list:
    \begin{enumerate}
        \item If the current interval does not overlap with the interval on the top of the stack, push it onto the top.
        \item If the current interval overlaps with stack top and ending time of current interval is more than that of stack top, update stack top with the ending time of current interval.
    \end{enumerate}
    \item Return the new list of intervals $[a_1,b_1],[a_2,b_2],\hdots,[a_M,b_M],\;M\leq N$.
\end{enumerate}
So, we can write the pseudocode for the approach as follows:\\
We are given a satellite with semimajor axis $a$ and orbit inclination $i$, and a set of ground stations with latitude-longitude coordinates $(\lambda_1,L_1),(\lambda_2,L_2),\hdots,(\lambda_N,L_N)$ and ground station mask radii $\theta_1,\theta_2,\hdots,\theta_N$.\\
Then, for each $\lambda$, we can write down the interval giving the range of values of $L$ for each ground station:
$$[L_{1,\min}(\lambda),L_{1,\max}(\lambda)],\;[L_{2,\min}(\lambda),L_{2,\max}(\lambda)],\hdots,[L_{N,\min}(\lambda),L_{N,\max}(\lambda)]$$
Now, there are two steps of preprocessing we need to do:
\begin{enumerate}
    \item Remove all empty intervals (ie. intervals corresponding to ground stations which have no visibility at the latitude $\lambda$.
    \item Some intervals may wrap around past $\pi$. Break these up into two intervals, wrapping the interval around:
    $$[L_{k,\min}(\lambda),L_{k,\max}(\lambda)]\rightarrow [L_{k,\min}(\lambda),\pi],\;[-\pi,L_{k,\max}(\lambda)-2\pi]$$
    (also do the analogous for intervals wrapping past $-\pi$)
\end{enumerate}
Label this new set of intervals: $[x_1,y_1],[x_2,y_2],\hdots,[x_{N'},y_{N'}]$
Then run the interval merge algorithm on these to yield the new list of ordered disjoint intervals:
$$I(\lambda):=\{[a_1,b_1],[a_2,b_2],\hdots,[a_M,b_M]\}$$
Furthermore, let:\\
$\lambda_{\min}=\trunc_i(\min\limits_k\{\lambda_k-\theta_k\})$,\\
$\lambda_{\max}=\trunc_i(\min\limits_k\{\lambda_k+\theta_k\})$.\\
Then, we can write the view period integral as:
\begin{align*}
\rho&=\int_{\lambda=\lambda_{\min}}^{\lambda_{\max}}\sum_{[a_k,b_k]\in I(\lambda)}\int_{L=a_k}^{b_k}\dfrac{\cos(\lambda)}{2\pi^2\sqrt{\sin^2(i)-\sin^2(\lambda)}}dLd\lambda\\
&=\int_{\lambda=\lambda_{\min}}^{\lambda_{\max}}\sum_{[a_k,b_k]\in I(\lambda)}\dfrac{\cos(\lambda)(b_k-a_k)}{2\pi^2\sqrt{\sin^2(i)-\sin^2(\lambda)}}d\lambda\\
&=\int_{\lambda=\lambda_{\min}}^{\lambda_{\max}}\dfrac{\cos(\lambda)}{2\pi^2\sqrt{\sin^2(i)-\sin^2(\lambda)}}\left(\sum_{[a_k,b_k]\in I(\lambda)}(b_k-a_k)\right)d\lambda\\
&=\dfrac{1}{2\pi^2}\int_{\lambda=\lambda_{\min}}^{\lambda_{\max}}\dfrac{\cos(\lambda)\text{Length}(I(\lambda))}{\sqrt{\sin^2(i)-\sin^2(\lambda)}}d\lambda
\end{align*}
(where $\text{Length}(I(\lambda)):=\sum_{[a_k,b_k]\in I(\lambda)}(b_k-a_k)$, the total length of the union of the intervals)\\\\
Finally, we take the change of variables from Equation \ref{eqn:LambdaChangeOfVariables} to remove any singularities from the integrand, yielding Equation \ref{eqn:CircularMultipleGSVPR}.

\bibliographystyle{AAS_publication}   
\bibliography{references}   

\end{document}